\documentclass{aastex62}
\usepackage{natbib}
\usepackage{amsmath}
\usepackage{graphicx}
\usepackage{bm}
\usepackage{rotating}

\newcommand\thetaE{\theta_\mathrm{E}}

\newcommand\piE{\pi_\mathrm{E}}

\received{June 1, 2020}
\revised{June 7, 2020}
\accepted{\today}
\submitjournal{AJ}

\shorttitle{MOA-2019-BLG-008}
\shortauthors{ROME/REA Team}
\begin{document}

\title{MOA-2019-BLG-008Lb: a new microlensing detection of an object at the planet/brown dwarf boundary.}

\correspondingauthor{E.~Bachelet}
\email{etibachelet@gmail.com}

\author[0000-0002-6578-5078]{E.~Bachelet}
\affiliation{Las Cumbres Observatory, 6740 Cortona Drive, Suite 102, Goleta, CA 93117, USA.}

\author[0000-0001-8411-351X]{Y.~Tsapras}
\affiliation{Zentrum f{\"u}r Astronomie der Universit{\"a}t Heidelberg, Astronomisches Rechen-Institut, M{\"o}nchhofstr. 12-14, 69120 Heidelberg, Germany}

\author{Andrew Gould}
\affiliation{Max-Planck-Institute for Astronomy, K\"onigstuhl 17, 69117 Heidelberg, Germany}
\affiliation{Department of Astronomy, Ohio State University, 140 W. 18th Ave., Columbus, OH 43210, USA}

\author{R.A.~Street}
\affiliation{Las Cumbres Observatory, 6740 Cortona Drive, Suite 102, Goleta, CA 93117, USA.}

\author{David P.~Bennett}
\affiliation{Code 667, NASA Goddard Space Flight Center, Greenbelt, MD 20771, USA}

\collaboration{and}

\author{M.P.G.~Hundertmark}
\affiliation{Zentrum f{\"u}r Astronomie der Universit{\"a}t Heidelberg, Astronomisches Rechen-Institut, M{\"o}nchhofstr. 12-14, 69120 Heidelberg, Germany}
%0000-0003-0961-5231

\author{V.~Bozza}
\affiliation{Dipartimento di Fisica "E.R. Canianiello", Universit{\`a} di Salerno, Via Giovanni Paolo II 132, 84084, Fisciano, Italy\\Istituto Nazionale di Fisica Nucleare, Sezione di Napoli, Via Cintia, 80126, Napoli, Italy}

\author{D.M.~Bramich}
\affiliation{New York University Abu Dhabi, Saadiyat Island, Abu Dhabi, PO Box 129188, United Arab Emirates}

\author{A.~Cassan}
\affiliation{Institut d’Astrophysique de Paris, Sorbonne Universit\'e, CNRS, UMR 7095, 98 bis bd Arago, 75014 Paris, France}

\author{M.~Dominik}
\affiliation{Centre for Exoplanet Science, SUPA, School of Physics \& Astronomy, University of St Andrews, North Haugh, St Andrews KY16 9SS, UK}

\author{K. Horne}
\affiliation{Centre for Exoplanet Science, SUPA, School of Physics \& Astronomy, University of St Andrews, North Haugh, St Andrews KY16 9SS, UK}

\author{S.~Mao}
\affiliation{Physics Department and Tsinghua Centre for Astrophysics, Tsinghua University, Beijing 100084, China}
\affiliation{National Astronomical Observatories, Chinese Academy of Sciences, 20A Datun Road, Chaoyang District, Beijing 100012, China}

\author{A.~Saha}
\affiliation{National Optical Astronomy Observatory, 950 North Cherry Ave., Tucson, AZ 85719, USA}

\author{J.~Wambsganss}
\affiliation{Zentrum f{\"u}r Astronomie der Universit{\"a}t Heidelberg, Astronomisches Rechen-Institut, M{\"o}nchhofstr. 12-14, 69120 Heidelberg, Germany}
\affiliation{International Space Science Institute (ISSI), Hallerstra{\ss}e 6, 3012 Bern, Switzerland}

\author[0000-0001-6000-3463]{Weicheng~Zang}
\affiliation{Physics Department and Tsinghua Centre for Astrophysics, Tsinghua University, Beijing 100084, China}

\collaboration{(The ROME/REA Collaboration)}

\author{Fumio Abe}
\affiliation{Institute for Space-Earth Environmental Research, Nagoya University, Nagoya 464-8601, Japan}
\author{Richard Barry}
\affiliation{Code 667, NASA Goddard Space Flight Center, Greenbelt, MD 20771, USA}
\author{David P.~Bennett}
\affiliation{Code 667, NASA Goddard Space Flight Center, Greenbelt, MD 20771, USA}
\affiliation{Department of Astronomy, University of Maryland, College Park, MD 20742, USA}
\author{Aparna Bhattacharya}
\affiliation{Code 667, NASA Goddard Space Flight Center, Greenbelt, MD 20771, USA}
\affiliation{Department of Astronomy, University of Maryland, College Park, MD 20742, USA}
\author{Ian A. Bond}
\affiliation{Institute of Natural and Mathematical Sciences, Massey University, Auckland 0745, New Zealand}
\author{Akihiko Fukui}
\affiliation{Department of Earth and Planetary Science, Graduate School of Science, The University of Tokyo, 7-3-1 Hongo, Bunkyo-ku, Tokyo 113-0033, Japan}
\affiliation{Instituto de Astrof\'isica de Canarias, V\'ia L\'actea s/n, E-38205 La Laguna, Tenerife, Spain}
\author{Hirosane Fujii}
\affiliation{Institute for Space-Earth Environmental Research, Nagoya University, Nagoya 464-8601, Japan}
\author{Yuki Hirao}
\affiliation{Department of Earth and Space Science, Graduate School of Science, Osaka University, Toyonaka, Osaka 560-0043, Japan}
\author{Yoshitaka Itow}
\affiliation{Institute for Space-Earth Environmental Research, Nagoya University, Nagoya 464-8601, Japan}
\author{Rintaro Kirikawa}
\affiliation{Department of Earth and Space Science, Graduate School of Science, Osaka University, Toyonaka, Osaka 560-0043, Japan}
author{Iona Kondo}
\affiliation{Department of Earth and Space Science, Graduate School of Science, Osaka University, Toyonaka, Osaka 560-0043, Japan}
\author{Naoki Koshimoto}
\affiliation{Department of Astronomy, Graduate School of Science, The University of Tokyo, 7-3-1 Hongo, Bunkyo-ku, Tokyo 113-0033, Japan}
\author{Yutaka Matsubara}
\affiliation{Institute for Space-Earth Environmental Research, Nagoya University, Nagoya 464-8601, Japan}
\author{Sho Matsumoto}
\affiliation{Department of Earth and Space Science, Graduate School of Science, Osaka University, Toyonaka, Osaka 560-0043, Japan}
\author{Shota Miyazaki}
\affiliation{Department of Earth and Space Science, Graduate School of Science, Osaka University, Toyonaka, Osaka 560-0043, Japan}
\author{Yasushi Muraki}
\affiliation{Institute for Space-Earth Environmental Research, Nagoya University, Nagoya 464-8601, Japan}
\author{Greg Olmschenk}
\affiliation{Code 667, NASA Goddard Space Flight Center, Greenbelt, MD 20771, USA}
\author{Cl\'ement Ranc}
\affiliation{Sorbonne Universit\'e, CNRS, UMR 7095, Institut d'Astrophysique de Paris, 98 bis bd Arago, 75014 Paris, France}
\author{Arisa Okamura}
\affiliation{Department of Earth and Space Science, Graduate School of Science, Osaka University, Toyonaka, Osaka 560-0043, Japan}
\author{Nicholas J. Rattenbury}
\affiliation{Department of Physics, University of Auckland, Private Bag 92019, Auckland, New Zealand}
\author{Yuki Satoh}
\affiliation{Department of Earth and Space Science, Graduate School of Science, Osaka University, Toyonaka, Osaka 560-0043, Japan}
\author{Takahiro Sumi}
\affiliation{Department of Earth and Space Science, Graduate School of Science, Osaka University, Toyonaka, Osaka 560-0043, Japan}
\author{Daisuke Suzuki}
\affiliation{Department of Earth and Space Science, Graduate School of Science, Osaka University, Toyonaka, Osaka 560-0043, Japan}
\author{Stela Ishitani Silva}
\affiliation{Department of Physics, The Catholic University of America, Washington, DC 20064, USA}
\affiliation{Code 667, NASA Goddard Space Flight Center, Greenbelt, MD 20771, USA}
\author{Taiga Toda}
\affiliation{Department of Earth and Space Science, Graduate School of Science, Osaka University, Toyonaka, Osaka 560-0043, Japan}
\author{Paul . J. Tristram}
\affiliation{University of Canterbury Mt.\ John Observatory, P.O. Box 56, Lake Tekapo 8770, New Zealand}
\author{Aikaterini Vandorou}
\affiliation{Code 667, NASA Goddard Space Flight Center, Greenbelt, MD 20771, USA}
\affiliation{Department of Astronomy, University of Maryland, College Park, MD 20742, USA}
\author{Hibiki Yama}
\affiliation{Department of Earth and Space Science, Graduate School of Science, Osaka University, Toyonaka, Osaka 560-0043, Japan}

\collaboration{(The MOA collaboration)}

\author{Michael D. Albrow}
\affiliation{University of Canterbury, Department of Physics and Astronomy, Private Bag 4800, Christchurch 8020, New Zealand}

\author{Sun-Ju Chung}
\affiliation{Korea Astronomy and Space Science Institute, Daejon 34055, Republic of Korea}

\author{Cheongho Han}
\affiliation{Department of Physics, Chungbuk National University, Cheongju 28644, Republic of Korea}

\author{Kyu-Ha Hwang}
\affiliation{Korea Astronomy and Space Science Institute, Daejon 34055, Republic of Korea}

\author{Youn Kil Jung}
\affiliation{Korea Astronomy and Space Science Institute, Daejon 34055, Republic of Korea}

\author{Yoon-Hyun Ryu}
\affiliation{Korea Astronomy and Space Science Institute, Daejon 34055, Republic of Korea}

\author{In-Gu Shin}
\affiliation{Korea Astronomy and Space Science Institute, Daejon 34055, Republic of Korea}

\author{Yossi Shvartzvald}
\affiliation{Department of Particle Physics and Astrophysics, Weizmann Institute of Science, Rehovot 76100, Israel}

\author{Jennifer C. Yee}
\affiliation{Center for Astrophysics $|$ Harvard \& Smithsonian, 60 Garden St.,Cambridge, MA 02138, USA}

\author{Sang-Mok Cha}
\affiliation{Korea Astronomy and Space Science Institute, Daejon 34055, Republic of Korea}
\affiliation{School of Space Research, Kyung Hee University, Yongin, Kyeonggi 17104, Republic of Korea} 

\author{Dong-Jin Kim}
\affiliation{Korea Astronomy and Space Science Institute, Daejon 34055, Republic of Korea}

\author{Seung-Lee Kim}
\affiliation{Korea Astronomy and Space Science Institute, Daejon 34055, Republic of Korea}

\author{Chung-Uk Lee}
\affiliation{Korea Astronomy and Space Science Institute, Daejon 34055, Republic of Korea}

\author{Dong-Joo Lee}
\affiliation{Korea Astronomy and Space Science Institute, Daejon 34055, Republic of Korea}

\author{Yongseok Lee}
\affiliation{Korea Astronomy and Space Science Institute, Daejon 34055, Republic of Korea}
\affiliation{School of Space Research, Kyung Hee University, Yongin, Kyeonggi 17104, Republic of Korea}

\author{Byeong-Gon Park}
\affiliation{Korea Astronomy and Space Science Institute, Daejon 34055, Republic of Korea}
\affiliation{University of Science and Technology, Korea, (UST), 217 Gajeong-ro Yuseong-gu, Daejeon 34113, Republic of Korea}

\author{Richard W. Pogge}
\affiliation{Department of Astronomy, Ohio State University, 140 W. 18th Ave., Columbus, OH  43210, USA}

\collaboration{(The KMTNet Collaboration)}

\author [0000-0001-5207-5619]{Andrzej Udalski}
\affiliation{Astronomical Observatory, University of Warsaw,
Al.~Ujazdowskie~4, 00-478~Warszawa, Poland}

\author[0000-0001-7016-1692]{Przemek Mr{\'o}z}
\affiliation{Astronomical Observatory, University of Warsaw,
Al.~Ujazdowskie~4, 00-478~Warszawa, Poland}

\author[0000-0002-9245-6368]{Rados{\l}aw Poleski}
\affiliation{Astronomical Observatory, University of Warsaw,
Al.~Ujazdowskie~4, 00-478~Warszawa, Poland}

\author[0000-0002-2335-1730]{Jan Skowron}
\affiliation{Astronomical Observatory, University of Warsaw,
Al.~Ujazdowskie~4, 00-478~Warszawa, Poland}

\author[0000-0002-0548-8995]{Micha{\l} K. Szyma{\'n}ski}
\affiliation{Astronomical Observatory, University of Warsaw,
Al.~Ujazdowskie~4, 00-478~Warszawa, Poland}

\author[0000-0002-7777-0842]{Igor Soszy{\'n}ski}
\affiliation{Astronomical Observatory, University of Warsaw,
Al.~Ujazdowskie~4, 00-478~Warszawa, Poland}

\author [0000-0002-2339-5899]{Pawe{\l} Pietrukowicz}
\affiliation{Astronomical Observatory, University of Warsaw,
Al.~Ujazdowskie~4, 00-478~Warszawa, Poland}

\author [0000-0003-4084-880X]{Szymon Koz{\l}owski}
\affiliation{Astronomical Observatory, University of Warsaw,
Al.~Ujazdowskie~4, 00-478~Warszawa, Poland}

\author[0000-0001-6364-408X]{Krzysztof Ulaczyk}
\affiliation{Department of Physics, University of Warwick, Gibbet Hill
Road, Coventry, CV4~7AL,~UK}

\author [0000-0002-9326-9329]{Krzysztof A. Rybicki}
\affiliation{Astronomical Observatory, University of Warsaw,
Al.~Ujazdowskie~4, 00-478~Warszawa, Poland}

\author[0000-0002-6212-7221]{Patryk  Iwanek}
\affiliation{Astronomical Observatory, University of Warsaw,
Al.~Ujazdowskie~4, 00-478~Warszawa, Poland}

\author[0000-0002-3051-274X]{Marcin Wrona }
\affiliation{Astronomical Observatory, University of Warsaw,
Al.~Ujazdowskie~4, 00-478~Warszawa, Poland}

\author[0000-0002-1650-1518]{Mariusz Gromadzki}
\affiliation{Astronomical Observatory, University of Warsaw,
Al.~Ujazdowskie~4, 00-478~Warszawa, Poland}

\collaboration{(The OGLE Collaboration)}

\begin{abstract}
We report on the observations, analysis and interpretation of the microlensing event MOA-2019-BLG-008. The observed anomaly in the photometric light curve is best described through a binary lens model. In this model, the source did not cross caustics and no finite source effects were observed. Therefore the angular Einstein ring radius $\thetaE$ cannot be measured from the light curve alone. However, the large event duration, $t_{\rm{E}}\sim80$ days, allows a precise measurement of the microlensing parallax $\piE$. In addition to the constraints on the angular radius $\theta_*$ and the apparent brightness $I_s$ of the source, we employ the Besan\c{c}on and GalMod galactic models to estimate the physical properties of the lens. We find excellent agreement between the predictions of the two Galactic models: the companion is likely a resident of the brown dwarf desert with a mass $M_p\sim30~M_{Jup}$ and the host is a main sequence dwarf star. The lens lies along the line of sight to the Galactic Bulge, at a distance of $\le 4$ kpc. We estimate that in about 10 years, the lens and source will be separated by $\sim55$ mas, and it will be possible to confirm the exact nature of the lensing system by using high-resolution imaging from ground or space-based observatories.
\end{abstract}

%% Keywords should appear after the \end{abstract} command. 
%% See the online documentation for the full list of available subject
%% keywords and the rules for their use.
\keywords{microlensing}

%% From the front matter, we move on to the body of the paper.
%% Sections are demarcated by \section and \subsection, respectively.
%% Observe the use of the LaTeX \label
%% command after the \subsection to give a symbolic KEY to the
%% subsection for cross-referencing in a \ref command.
%% You can use LaTeX's \ref and \label commands to keep track of
%% cross-references to sections, equations, tables, and figures.
%% That way, if you change the order of any elements, LaTeX will
%% automatically renumber them.
%%
%% We recommend that authors also use the natbib \citep
%% and \citet commands to identify citations.  The citations are
%% tied to the reference list via symbolic KEYs. The KEY corresponds
%% to the KEY in the \bibitem in the reference list below. 

\section{Introduction} 
\label{sec:intro}
During the last 20 years, thousands of planets\footnote{4940 to date according to \url{https://exoplanetarchive.ipac.caltech.edu/}} have been detected and it is now clear that planets are abundant in the Milky Way \citep{Cassan2012,Bonfils2013,Clanton2016,Fulton2021}. Conversely, the various methods of detection agree that brown dwarf companions (with a mass $\sim$13-80 $M_{Jup}$) seem much rarer \citep{Grether2006,Lafreniere2007,Kraus2008,Metchev2009,Kiefer2019,Nielsen2019,Carmichael2020}, inspiring the idea of a ``brown dwarf desert" \citep{Marcy2000}, and such disparity raises questions about formation scenarios. Core accretion, disc instability, migration and disc evolution mechanisms are capable of producing planets up to $\lesssim40 M_{Jup}$ \citep{Pollack1996,Boss1997,Alibert2005,Mordasini2009}, explaining the formation of some brown dwarf companions. Brown dwarfs can also form via gas-collapse \citep{Bejar2001,Bate2002} and several processes have been proposed to explain the cessation of gas accretion, such as ejection (see \citet{Luhman2012} and references therein for a more complete review). However, the formation of low-mass binaries remains difficult to explain \citep{Bate2002,Marks2017} and more detections are needed to place meaningful constraints on formation models, especially around the brown dwarf desert.

Several objects at the planet/brown dwarf mass boundary have been discovered with the microlensing technique, both in binary and single lens events \citep{Bachelet2012b,Bozza2012,Ranc2015,Han2016,Zhu2016,Poleski2017,Shvartzvald2019,Bachelet2019,Miyazaki2020}.
Microlensing is particularly sensitive to exoplanets and brown dwarfs at or beyond the snow-line of their host stars, which is the region beyond which it is cold enough for water to turn to ice. Planets in this region typically have orbital periods of many years and, as such, are mostly inaccessible to other planet detection methods \citep{Gould2010,Tsapras2016}. The location of the snow-line plays an important role during planet formation, as the prevalence of ice grains beyond that point is believed to facilitate the formation of sufficiently large planetary cores, able to trigger runaway growth and form giant planets \citep{Ida2004,Kley2012}.

The lensing geometry is typically expressed in terms of the angular Einstein radius of the lens \citep{Einstein1936}
\begin{equation}
    \thetaE = \sqrt{\frac{D_\mathrm{LS}}{D_\mathrm{S} D_\mathrm{L}}\frac{4GM_\mathrm{L}}{c^2}},
\end{equation}
where $D_\mathrm{L}, D_\mathrm{S}$ are the distances from the observer to the lens and source respectively, $D_\mathrm{LS}$ is the lens-source distance, and $M_\mathrm{L}$ is the mass of the lens. The key observable in microlensing events that provides any connection to the physical properties of the lens is the event timescale
\begin{equation}
    t_E = \frac{\thetaE}{\mu_{\mathrm{rel}}}=\frac{\sqrt{\kappa M_L \pi_{\mathrm{rel}}}}{\mu_{\mathrm{rel}}},
\end{equation}
where $\mu_{\mathrm{rel}}$ is the relative proper motion between lens and source, $\pi_{\mathrm{rel}}$ is the lens-source relative parallax and $\kappa\equiv 4G/c^2\mathrm{au}\simeq8.1\mathrm{mas}/M_\odot$ \citep{Gould2000}. These two equations reveal a well-known degeneracy in microlensing event parameters. Indeed, the mass of the lens, its distance and the distance to the source are degenerate parameters when only $t_{\rm{E}}$ is measured and at least two extra pieces of information are required to disentangle them. For binary lenses, $\thetaE=\theta_*/\rho$ is often measured from the detection of finite source effects in the event light curve, typically parametrized with $\rho$. This occurs when an extended source of angular radius $\theta_*$ closely approaches regions of strong magnification gradients, i.e. around caustics \citep{Witt1990, Tsapras2018}. Using a color-radius relation \citep{Boyajian2014}, it is then possible to estimate $\thetaE$. For sufficiently long events (i.e., $t_E\ge30$ days), the microlensing parallax $\piE=\pi_{rel}/\thetaE$ due to Earth's revolution around the Sun, can be measured. This is referred to as the annual parallax \citep{Gould1992,Gould2004}. In addition, if simultaneous observations can be performed from space, as well as from the ground, it is possible to measure the space-based parallax \citep{Refsdal1966,Novati2015,Yee2015b}. Ultimately, by obtaining high-resolution imaging several years after the event has expired, additional constraints on the relative lens-source proper motion $\mu_{rel}$ and the lens brightness may be obtained, provided that the lens and source can be resolved \citep{Alcock2001}. See for example \citet{Beaulieu2018b} for a complete review of this technique.

It is not rare however that only $t_{\rm{E}}$ and a single other parameter ($\thetaE$ or $\piE$) are measured, leaving the physical parameters of the lens system only loosely constrained. As underlined by \citet{Penny2016}, this is the case for about 50 \% of all published microlensing planetary events. To obtain stronger constraints on these events, Galactic models may be employed to derive the probability densities of lens mass and distance along the line of sight that reproduce the fitted microlensing event parameters. Originally used by \citet{Han1995}, Galactic models are now commonly relied upon to estimate the properties of microlensing planets when no additional information is available to constrain the parameter space \citep{Penny2016}. While they  generally come with large uncertainties (10\% is a lower limit), Galactic model predictions have proven to be in excellent agreement with results obtained from follow-up studies using high-resolution imaging, especially for OGLE-2005-BLG-169Lb \citep{Gould2006,Batista2015,Bennett2015}, MOA-2011-BLG-293Lb \citep{Yee2012,Batista2014} and OGLE-2014-BLG-0124Lb \citep{Udalski2015,Beaulieu2018}. Galactic models developed for microlensing analysis are employed to generate distributions of stellar densities and velocities across the Galactic Disk and Bulge \citep{Han1995,Han2003,Dominik2006,Bennett2014}, and use them to reproduce microlensing observables (i.e. $t_{\rm{E}}$, $\thetaE$ and $\piE$). These are then compared to the fitted event parameters in order to estimate the probability densities of the lens distance $D_l$ and mass $M_l$. In addition to these models, there exist synthetic stellar population models for the Milky Way that have been explicitly developed to reproduce observable Galactic properties with great accuracy. Specifically, the Besan\c{c}on \citep{Robin2003} and GalMod \citep{pasetto2018a} models have been used in many different studies, to explore the structure, kinematics and formation history of the the Milky Way \citep{Czekaj2014, Robin2017}. In addition, they have also been used to simulate astronomical sky-surveys \citep{Rauer2014,Penny2013,Penny2019,Kauffman2020}, and their predictions have been tested against real observations \citep{Schultheis2006,Bochanski2007,Pietrukowicz2012,Schmidt2020,Terry2020}.

% You might say something more here on the evolution of Galactic models used in microlensing (Dominik, Bennett, Gould, Besancon, ...)

For the first time, in this study we employ both the Besan\c{c}on and GalMod Galactic models to estimate the properties of a binary lens, with a companion likely located in the brown dwarf desert. The microlensing event MOA-2019-BLG-008 was observed by several microlensing teams independently, and we present the different data sets, as well as the data reduction procedures, in Section~\ref{sec:obs}. The modeling of the photometric light curve and the model selection are discussed in Section~\ref{sec:modeling}. Section~\ref{sec:color_analysis} presents the analysis of the properties of the source and of the blend contaminant. The methodology used to derive the physical properties of the lens system is detailed in Section~\ref{sec:galmod}, and we conclude in Section~\ref{sec:concl}.  

\section{Observations and Data Reduction} 
\label{sec:obs}

\subsection{Survey and follow-up observations}
The microlensing event MOA-2019-BLG-008 was first announced on 4 Feb 2019 by the MOA collaboration \citep{sumi2003}, which operates the 1.8-m MOA survey telescope at Mount John observatory in New Zealand, at equatorial coordinates $\alpha = 17\textsuperscript{h}51\textsuperscript{m}55.89\textsuperscript{s}$, $\delta = -29^{\circ}59\arcmin23.03\arcsec$ (J2000) ($l,b = 359.8049^{\circ}, -1.7203^{\circ}$). The event was also independently identified by the Early Warning System (EWS)\footnote{http://ogle.astrouw.edu.pl/ogle4/ews/ews.html} of the Optical Gravitational Lensing Experiment (OGLE) survey \citep{Udalski2003,Udalski2015} as OGLE-2019-BLG-0011. OGLE observations were carried out with the 1.3-m Warsaw telescope at Las Campanas Observatory in Chile, with the 32-chip mosaic CCD camera. The event occurred in OGLE bulge field BLG501, which was imaged about once per hour when not interrupted by weather or the full Moon, providing good coverage of the light curve when the bulge was visible from Chile.

Additional observations were obtained by the ROME/REA survey \citep{Tsapras2019} using 6$\times$1m telescopes from the southern ring of the global robotic telescope network of the Las Cumbres Observatory (LCO) \citep{brown2013}. The LCO telescopes are located at the Cerro Tololo International Observatory (CTIO) in Chile, South African Astronomical Observatory (SAAO) in South Africa and Siding Spring Observatory (SSO) in Australia, and they provided good coverage of the light curve, although the event occurred early in the 2019 ROME/REA microlensing season (i.e. $\sim$ March to September of each year, when the Galactic Bulge is observable). Obsevrations were acquired in the survey mode.

The event lies in fields BLG02 and BLG42 of the Korea Microlensing
Telescopes Network (KMTNet) \citep{kim2016} and so, was intensely
monitored by that survey, although KMTNet did not independently
discover the event. Observations were also obtained from the Spitzer
satellite as part of an effort to constrain the parallax \citep{yee2015}. Spitzer observations will be presented in a companion paper (Han et al. 2022 in prep.).
    
\subsection{Data reduction procedure}
This analysis uses all available ground-based observations of MOA-2019-BLG-008. The list of contributing telescopes is given in Table \ref{tab:telescopes2}. Most data were obtained in the $I$ band (or SDSS-i) but we note that MOA observations were performed with the MOA wide-band red filter, which is specific to that survey \citep{sako2008}. ROME/REA obtained observations in three different bands (SDSS-i$^{\prime}$, SDSS-r$^{\prime}$ and SDSS-g$^{\prime}$). The KMTNet survey observations were carried out in the $I$ band, with a complementary $V$ band observation every ten $I$ exposures.

The photometric analysis of crowded-field observations is a challenging task. Images of the Galactic bulge contain hundred of thousands of stars whose point-spread functions (PSFs) often overlap, therefore aperture and PSF-fitting photometry offer very limited sensitivity to photometric deviations generated by the presence of low-mass planetary companions. For this reason, observers of microlensing events routinely perform difference image analysis (DIA) \citep{Tomaney1996,1998ApJ...503..325A,2008MNRAS.386L..77B,2013MNRAS.428.2275B}, which offers superior photometric precision under such crowded conditions. 
%For a given telescope and camera, the technique of difference image analysis uses a {\it reference} image\footnote{This can be either a single image or a combination of images taken under the best seeing conditions} to which background, astrometric, photometric and point-spread-function corrections are applied to match the images of that same field taken at each individual epoch. The fitted model based on the {\it reference} image is then subtracted from the matching images to produce residual (or {\it difference} images). Stars that did not vary in brightness between the times the images were obtained leave no systematic residuals on the {\it difference} images, but stars that underwent brightness variations leave clear positive or negative residuals. 
Most microlensing teams have developed custom DIA pipelines to reduce their observations. OGLE, MOA and KMT images were reduced using the photometric pipelines described in \citet{Udalski2003}, \citet{Bond2001} and \citet{Albrow2009}, respectively. The LCO %and MiNDSTEp 
observations were processed using the \textit{pyDANDIA} pipeline (ROME/REA in prep), a customized re-implementation of the \textit{DanDIA} pipeline \citep{Bramich2008,Bramich2013} in Python. The data sets presented in this paper have been carefully reprocessed to achieve greater photometric accuracy, and it is these data that we used as input when modelling the microlensing event. They are available for download from the online version of the paper. %These data sets are released together with the paper and are available for download.

\begin{figure}[h]
    \centering
    \includegraphics[width=0.45\columnwidth]{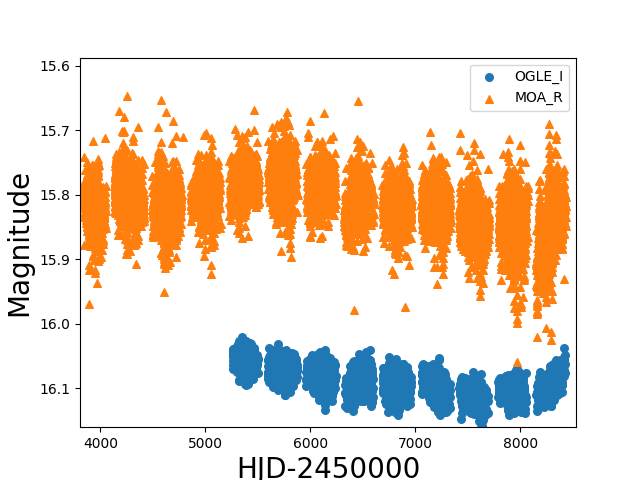}
    \caption{OGLE and MOA long-term baseline trends.}
    \label{fig:trend}
\end{figure}

%During years prior to the microlensing event peak,  we noticed a small trend 
We note the presence of a very long-term baseline trend spanning several observing seasons in the OGLE and MOA photometry that can be seen in Figure~\ref{fig:trend}. As described later in this work, we determined that the source star is a red giant. Many red giants exhibit variability at the $\sim10\%$ level \citep{Wray2004,Percy2008,Wyrzykowski2006,Soszynski2013,Arnold2020}, and it is possible that this is also the case for this source, despite the apparently very long period $P\ge1000$ days. Because this trend manifests over very long time scales (several years), much longer than the duration of the actual microlensing event (weeks), it does not have any noticeable effect on the determination of the parameters of this event, which are primarily derived from the detailed morphology of the microlensing light curve. The baseline over the duration of the microlensing event is effectively flat. Therefore, to increase the speed of the modeling process, we only used observations with $JD\ge2457800$ and included data sets with more than 10 measurements in total during the course of the microlensing event. The latter constraint applies only to the LCO data and is limited to the observations acquired by the reactive REA mode on a different target in the same field \citep{Tsapras2019}. We verified that our data selection does not impact the overall results, by exploring models with the full-baseline. 

\begin{table}[h!]
\centering
\caption{Summary of telescopes and observations used for modeling the event. The number of data points per telescope represents the points used for the modeling step., i.e. $JD\ge2457800$. Lines marked with `*' indicate that this data set was not used during the modeling process, as described in the text. In cases which the rescaling parameters were not constrained, they were fixed to $k=1.0$ and $e_{min} = 0.0$. }% What is the error on the rescaling k of the error? Is it used in any way? 
\label{tab:telescopes2}
\begin{tabular}{lllllll}
\hline
\hline
Name& Site         & Aperture(m) &  Filters & k&$e_{min}$ & $N_{data}$ \\
\hline
$OGLE\_I  $  & Chile        & 1.3   & I      & $1.07\pm0.02$     & 0.0  & 2257 \\

$MOA\_R   $   & New Zealand  & 2.0     & Red    & $1.39\pm0.06$ & $0.009\pm0.001$    & 7824 \\
$MOA\_V   $   & New Zealand  & 2.0     & V    & $1.11\pm0.05$ & $0.012\pm 0.004$ & 253 \\

$KMTC02\_I  $  & Chile        & 1.6   & I      & $1.07\pm0.05$     & $0.0075 \pm 0.0003$ & 1542 \\
$KMTC02\_V  $  & Chile        & 1.6   & V      & $0.89\pm0.08$     & $0.003 \pm 0.002$ & 119 \\

$KMTA02\_I $  & Australia    & 1.6   & I      & $1.01\pm0.05$     & $0.0084 \pm 0.0004$ & 1298 \\
$KMTA42\_I $  & Australia    & 1.6   & I      & $0.92\pm0.09$     & $0.0078 \pm 0.0004$ & 1391 \\

$KMTC42\_I $  & Chile        & 1.6   & I      & $0.89\pm0.04$     & $0.0088\pm 0.0002$ & 1730 \\

$KMTS02\_I $ & South Africa & 1.6   & I       & $1.11\pm0.04$     & $0.0076 \pm 0.0003$ & 1458 \\
$KMTS42\_I $  & South Africa & 1.6   & I      & $0.89\pm0.04$     & $0.0079 \pm 0.0002$ & 1522 \\

$LCO\_COJA_\_gp$ & Australia & 1.0   & SDSS-g$^{\prime}$      & $0.96\pm0.05$     & $0.020\pm0.004$ & 133\\
$LCO\_COJA_\_rp $& Australia & 1.0   & SDSS-r$^{\prime}$      & $0.87\pm0.04$     & $0.025\pm0.002$ & 194\\
$LCO\_COJA_\_ip $& Australia & 1.0   & SDSS-i$^{\prime}$      & $1.01\pm0.09$     & $0.030 \pm 0.005$ & 310\\

$LCO\_COJB_\_gp $& Australia & 1.0   & SDSS-g$^{\prime}$      & * & * & *\\
$LCO\_COJB_\_rp $& Australia & 1.0   & SDSS-r$^{\prime}$      & *& *& *\\
$LCO\_COJB_\_ip $& Australia & 1.0   & SDSS-i$^{\prime}$      & $0.8\pm0.1$     &$0.007\pm0.006$& 21\\

$LCO\_CPTA_\_gp $& South Africa & 1.0   & SDSS-g$^{\prime}$      & $1.2\pm0.1$     & $0.010\pm0.005$& 104\\
$LCO\_CPTA_\_rp $& South Africa & 1.0   & SDSS-r$^{\prime}$      & $1.08\pm0.03$     & $0.020\pm0.004$& 141\\
$LCO\_CPTA_\_ip $& South Africa & 1.0   & SDSS-i$^{\prime}$      & $1.03\pm0.10$     & $0.023\pm0.003$ & 167\\

$LCO\_CPTB_\_gp $& South Africa & 1.0   & SDSS-g$^{\prime}$      & *& * & *\\
$LCO\_CPTB_\_rp $& South Africa & 1.0   & SDSS-r$^{\prime}$      & *& *& *\\
$LCO\_CPTB_\_ip $& South Africa & 1.0   & SDSS-i$^{\prime}$      &*     & * & *\\

$LCO\_CPTC_\_gp $& South Africa & 1.0   & SDSS-g$^{\prime}$      & *  &* & *\\
$LCO\_CPTC_\_rp $& South Africa & 1.0   & SDSS-r$^{\prime}$      & * & * & *\\
$LCO\_CPTC_\_ip $& South Africa & 1.0   & SDSS-i$^{\prime}$      & *     & *& *\\

$LCO\_LSCA_\_gp $& Chile & 1.0   & SDSS-g$^{\prime}$      & $1.03\pm0.04$     & $0.018\pm0.010$ & 99\\
$LCO\_LSCA_\_rp $& Chile & 1.0   & SDSS-r$^{\prime}$      & $1.06\pm0.05$     & $0.020\pm0.004$ & 142\\
$LCO\_LSCA_\_ip $& Chile & 1.0   & SDSS-i$^{\prime}$      & $1.06\pm0.05$     & $0.023\pm0.003$ & 273\\
$LCO\_LSCB_\_ip$ & Chile & 1.0   & SDSS-i$^{\prime}$      & *    &* & *\\

\hline
\hline
\end{tabular}
\end{table}

\section{Modeling the event light curve}
\label{sec:modeling}
This event displays a clear anomaly around $HJD\sim2458580$, implying that it is most likely due to a binary lens (2L1S) or a binary source (1L2S) \citep{Dominik2019}. {It is morpholigicaly similar to the event MACHO 99-BLG-47 \citep{Albrow2002}, despite a different lensing geometry, as detailed below}. In addition, because the event lasts for $\sim 300$ days, the effect of the motion of Earth around the Sun, referred to as the annual parallax \citep{Gould1992,Alcock1995}, needs to be taken into account. The classical approach to modeling is to first search for static binary models and subsequently gradually introduce additional second-order effects, such as parallax or the orbital motion of the lens \citep{Dominik1999}. To model the event we use the pyLIMA software \citep{Bachelet2017}, which employs the VBBinaryLensing code \citep{Bozza2010,Bozza2018} to estimate the binary lens model magnification, and we search for a general solution including the annual parallax, but we also explore the static case for completeness. The first step of modeling involves identifying potential multiple minima in the parameter space, and for this we employ the differential evolution algorithm \citep{Storn1997,Bachelet2017}. During the modeling process, we rescale the data uncertainties using the same method presented in \citet{Bachelet2019}, which introduces the parameters $k$ and $e_{\rm min}$ for each datasets:
\begin{equation}
\sigma ' = \sqrt{k^2\sigma^2+e_{\rm min}^2}
\end{equation}
where $\sigma$ and $\sigma '$ are the original and rescaled uncertainties (in magnitude units), respectively. The coefficients for each dataset are given in Table~\ref{tab:telescopes2}.  Finally, we explore the posterior distribution of the parameters of each minimum that we identify using the emcee software \citep{Foreman2013}.

\subsection{(No) finite-source effects}
\label{sec:norho}
\begin{figure}[h]
    \centering
    \includegraphics[width=0.99\columnwidth]{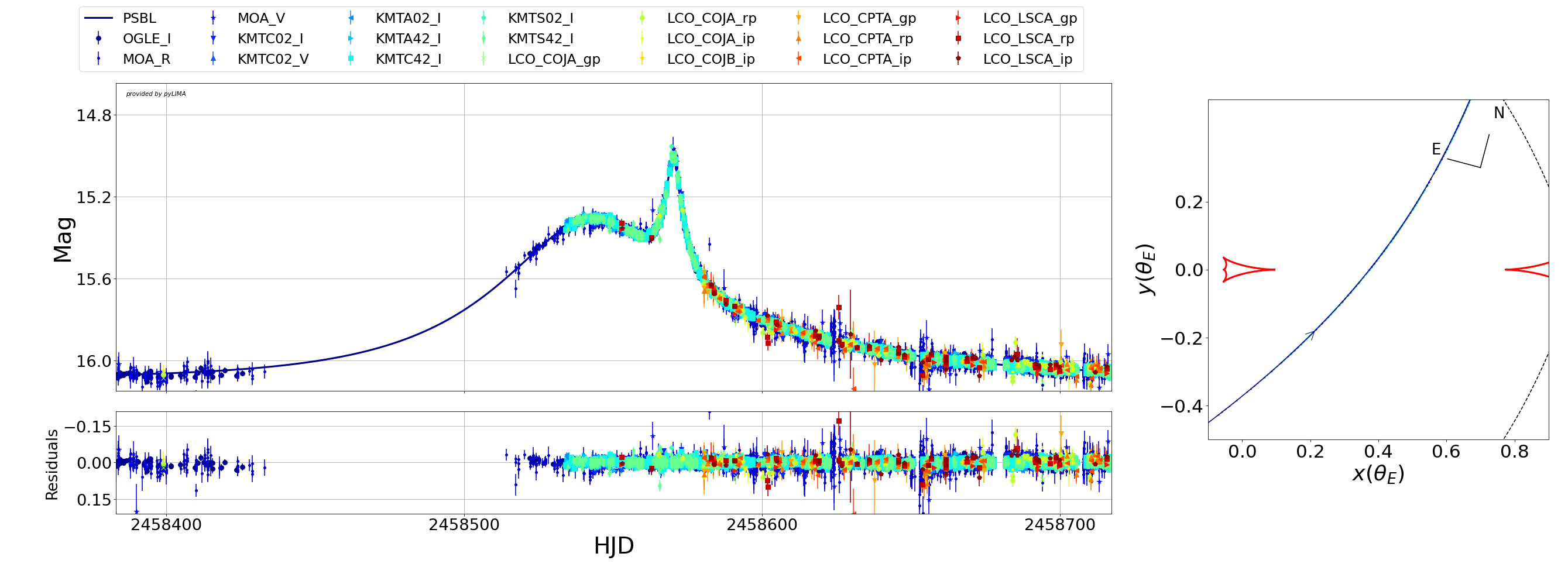}
    \caption{(Left) Observations and 2L1S-P model ($u_0\le0$ and no orbital motion) centered at the peak magnification. (Right) The corresponding model geometry, where red lines represent the caustics, dashed lines represent the critical curve, blue line is the source trajectory. The North and East vectors are also represented. The lens center of mass is fixed at (0,0).}
    \label{fig:model}
\end{figure}

In principle, the normalized angular source radius $\rho=\theta_*/\thetaE$ has to be considered \citep{Witt1994}, but preliminary models indicated that this parameter is loosely constrained. This is because the source trajectory does not pass close to caustics, as can be inferred from Figure~\ref{fig:model}. However, there exists an upper limit $\rho_\mathrm{m}$ where the finite-source effects would start to be significantly visible in the models. Because $\thetaE\ge\theta_*/\rho_m$, this limit introduces constraints on the mass and distance of the lens that can be used for the analysis presented in Section~\ref{sec:galmod}. Indeed, it is straightforward to derive \citep{Gould2000}:
\begin{equation}
\pi_l\ge{{\pi_E\theta_*}\over{\rho_m}}+\pi_s
\end{equation}
where $\pi_l$ and $\pi_s$ are the parallax of the lens and source, respectively.  The constraint on the mass can be written as
\begin{equation}
M_\mathrm{L}\ge\frac{\theta_*}{\kappa\piE\rho_\mathrm{m}}
\end{equation}
Therefore, we sample the distribution of $\rho $ around the best model and found that $\rho_m\le0.01$ (with a conservative $10\sigma$ limit) and consider the source as a point for the rest of the modeling presented in this analysis. 

\subsection{Binary lens}
 A binary lens model involves seven parameters. $t_0$ is the time when the angular distance $u_0$ (scaled to $\theta_\mathrm{E}$ between the source and the center of mass of the binary lens is minimal. The event duration is characterised by the angular Einstein ring radius crossing time $t_{\rm{E}}=\thetaE/\mu_{rel}$, where $\mu_{rel}$ is the lens/source relative proper motion (in the geocentric frame, because pyLIMA follows the geocentric formalism of \citealt{Gould2004}. The binary separation projected on the plane of the lens is defined as $s$ and the mass ratio between two component as $q$. The angle between the trajectory and the binary axis (fixed along the x-axis) is defined as $\alpha$. We also consider a source flux $f_s$ and a blend flux $f_b$ for each of the datasets, adding 2n parameters where n is the number of datasets (i.e., 29 in this study). As discussed previously, we neglect the last parameter $\rho$ and fit the simple point-source binary lens model. 
%in trigonometric system. 

Following \citet{Gould2004}, we define the parallax vector $\bm{\piE}$ by its North ($\pi_\mathrm{E,N}$) and East ($\pi_\mathrm{E,E}$) components. We set the parallax reference time as $t_{0,\mathrm{par}} = 2458570$ HJD \citep{Skowron2011} for all models considered in this analysis. This coincides with the time of the anomaly peak and corresponds to the calendar date 28 March 2019. At this time, Earth's acceleration vector was nearly parallel to the East direction. Therefore, we expect the $\pi_{EE}$ component to be the better constrained of the two. We found that the lightcurve morphology can only be explained by a single geometry, in agreement with previous results from real time modeling conducted by V. Bozza\footnote{http://www.fisica.unisa.it/gravitationAstrophysics/RTModel/2019/OB190011.htm} and C. Han\footnote{http://astroph.chungbuk.ac.kr/~cheongho/modelling/2019/FIG/OB-19-0011.jpg}. However, a second solution exists, a consequence of the binary ecliptic degeneracy \citep{Skowron2011} with $(u_0,\alpha,\pi_\mathrm{E,N}) \Longleftrightarrow -(u_0,\alpha,\pi_\mathrm{E,N})$. Because the magnification pattern is symmetric relative to the binary axis, it exists two source trajectories that produce identical lightcurves for static binaries, i.e. $(u_0,\alpha) \Longleftrightarrow -(u_0,\alpha)$. Moreover, the projected Earth acceleration can be considered as almost constant during the duration of the event, leading to the $\pi_{E,N}$ degeneracy for events located towards the Galactic Bulge \citep{Smith2003,Jiang2004,Gould2004,Poindexter2005,Skowron2011}. This degeneracy is especially severe for events occurring near the equinoxes, because the projected Earth acceleration varies slowly \citep{Skowron2011}. We therefore explore both solution in the following analysis. Note that pyLIMA uses the formalism of \citet{Gould2004} and therefore $u_0\ge0$ if the lens passes the source on its right. Given the relatively long timescale of the event, we also explored the possibility of orbital motion of the lens \citep{Albrow2000,2012ApJ...754...73B} and considered the simplest linear model, parametrized with $\mathrm{d}s/\mathrm{d}t$ and $\mathrm{d}\alpha/\mathrm{d}t$, the linear separation and angular variation over time at time $t_0$. For completeness, we explored the parameter space for a static model (i.e., without the annual parallax) and found that the best model converges to a similar geometry. However, the residuals display systematic trends around the event peak that are the clear signature of annual parallax, which is reflected in the high $\chi^2$ value presented in Table~\ref{tab:models}.

\subsection{Binary source}
We explored the possibility that the observed light curve was due to a binary source \citep{Gaudi1998}. Following the approach described in \citet{Hwang2013}, we added to the single lens model the extra parameters $\Delta t_0$ and $\Delta u_0$, respectively the shifts in the time of peak and separation of the second source relative to the first. Finally, we also added the flux ratio of the two sources in each observed band $q_\lambda$. We report our results in Table~\ref{tab:models}. We also explore models with two different source angular sizes, but did not find any significant improvements in the model likelihood.

\begin{table}[h!]
\footnotesize
\centering
\caption{Models parameters are defined as 16,50 and 84 MCMC chains percentile, except for the $\chi^2$ which is reported as the minimum value (i.e., the best model in each case). The angular source radius $\theta_*$ for each model is also presented, see Section~\ref{sec:color_analysis}. The two static models are denoted 2L1S, the two models with parallax are denoted 2L1SP and models with orbital motion of the lens are 2L1SPOM. + and - indicated positive or negative $u_0$.} \label{tab:models}
\begin{tabular}{llllllll}
\hline
\hline
Parameters [unit]&2L1S-& 2L1S+&2L1S-P & 2L1S+P&2L1S-POM & 2L1S+POM& 1L2S \\
\hline
$t_0$ [HJD-2450000] &$8546.47^{+0.06}_{-0.06}$ &$8546.47^{+0.06}_{-0.06}$&$8550.6^{+0.2}_{-0.3}$&$8547.8^{+0.2}_{-0.3}$ & $8547.5^{+0.4}_{-0.4}$ &$8546.7^{+0.2}_{-0.2}$& $8541.8^{+0.1}_{-0.1}$ \\
$u_0$ &$-0.287^{+0.002}_{-0.002}$&$0.287^{+0.002}_{-0.002}$& $-0.293^{+0.003}_{-0.003}$ & $0.294^{+0.003}_{-0.004}$ &$-0.360^{+0.005}_{-0.005}$&$0.354^{+0.005}_{-0.004}$&$0.195^{+0.005}_{-0.004}$ \\
$t_{\rm{E}}$ [days] & $93.4^{+0.4}_{-0.5}$&$93.4^{+0.5}_{-0.5}$&$83.7^{+0.5}_{-0.5}$& $87.9^{+0.7}_{-0.5}$&$79.2^{+0.7}_{-0.7}$& $80.9^{+0.6}_{-0.7}$ & $114^{+2}_{-2}$ \\
$\pi_{EN}$ &*&*& $-0.23^{+0.02}_{-0.02}$ & $0.07^{+0.01}_{-0.01}$&$-0.16^{+0.03}_{-0.03}$&$0.11^{+0.02}_{-0.02}$ & $0.109^{+0.007}_{-0.007}$ \\
$\pi_{EE}$ &*&*& $0.107^{+0.002}_{-0.002}$ &$0.110^{+0.003}_{-0.003}$ &$0.136^{+0.005}_{-0.005}$&$0.142^{+0.005}_{-0.004}$&$0.1232^{+0.006}_{-0.006}$ \\
$s $ & $1.5827^{+0.0009}_{-0.0009}$ &$1.5828^{+0.0009}_{-0.0009} $& $1.664^{+0.005}_{-0.006}$ & $1.620^{+0.004}_{-0.005}$ & $1.558^{+0.002}_{-0.002}$&$1.545^{+0.002}_{-0.002}$ &  * \\
$q $ & $0.0267^{+0.0001}_{-0.0001}$&$0.0267^{+0.0001}_{-0.0001}$&$0.0395^{+0.0009}_{-0.0011}$ & $0.0320^{+0.0007}_{-0.0008}$ &$0.0201^{+0.0004}_{-0.0004}$ & $0.0187^{+0.0003}_{-0.0003}$& * \\
$\alpha [rad]$ & $-2.297^{+0.001}_{-0.001}$&$2.297^{+0.001}_{-0.001}$&$-2.247^{+0.003}_{-0.003}$ &$2.285^{+0.003}_{-0.003}$ & $-2.243^{+0.005}_{-0.006}$& $2.257^{+0.003}_{-0.003}$ & *\\
$ds/dt [yr^{-1}] $ &*&*&*&* &$0.43^{+0.04}_{-0.04}$ & $0.47^{+0.03}_{-0.03}$ &* \\
$d\alpha/dt [yr^{-1}] $ &*&*&*&* &$-1.46^{+0.09}_{-0.09}$& $1.51 ^{+0.06}_{-0.07}$&* \\

$\Delta t_0 [days]$ &*&*&*&* &*&*&$28.2^{+0.1}_{-0.1}$ \\
$\Delta u_0$ &*&*&*&* &*&*&$-0.210^{+0.004}_{-0.005}$ \\
$q_V $ &*&*&*&* &*&*&$0.0910^{+0.006}_{-0.006}$ \\
$q_{\rm{MOA_R}} $ &*&*&*&* &*&*&$0.0916^{+0.001}_{-0.001}$ \\
$q_I $ &*&*&*&* &*&*&$0.0876^{+0.001}_{-0.001}$\\
$q_{gp} $ &*&*&*&* &*&*&$0.4^{+0.4}_{-0.3}$ \\
$q_{rp} $ &*&*&*&* &*&*&$0.11^{+0.03}_{-0.02}$ \\
$q_{ip} $ &*&*&*&* &*&*&$0.10^{+0.02}_{-0.01}$ \\

$\chi^2$ &{\bf 25336} &{\bf 25336} & {\bf 22556} & {\bf 22735} & {\bf 22351}& {\bf 22333} & {\bf 27203} \\
\hline
$\theta_*\rm{[\mu as]}$ & $10.4\pm0.3$ & $10.4\pm0.3$ & $10.5\pm0.3$ & $10.5\pm0.3$&$11.5\pm0.3$&$11.5\pm0.3$&*\\

\hline
\hline
\end{tabular}
\end{table}

\section{Analysis of the source and the blend}
\label{sec:color_analysis}
In the analysis of microlensing events, the color-magnitude diagram (CMD) is used to estimate the angular source radius $\theta_*$, and ultimately the angular Einstein ring radius $\thetaE=\theta_*/\rho$ \citep{Yoo2004}. Unfortunately, $\rho$ is not measurable in the present case. Nonetheless, the CMD analysis provides useful information about the source and the lens that can be used to place additional constraints to the analysis presented in Section~\ref{sec:galmod}. The CMD constraints can also be used to inform observing decisions in the future with complementary high-resolution imaging. We conducted our CMD analysis using different and independently obtained sets of observations from our pool of available data sets. The estimation of $\theta_*$ below is for the model 2L1S with parallax and $u_0<0$. The source and blend magnitudes for all models are presented in Table~\ref{tab:source_blend} and the angular source radius $\theta_*$ derived for all models is presented in Table~\ref{tab:models}.

\subsection{ROME/REA Color-Magnitude Diagrams Analysis}

\begin{figure}
    \centering
    \includegraphics[width=0.95\columnwidth]{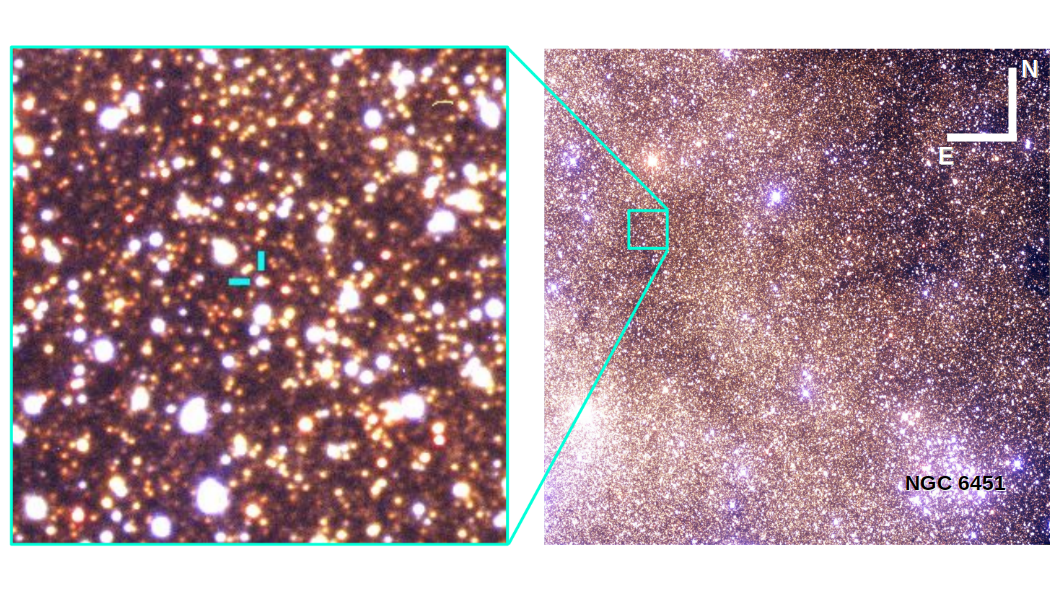}
    \caption{Color composite of g,r,i reference images of the ROME survey. The inset is 2'x2' zoom around MOA-2019-BLG-008. The NGC6451 cluster center is visible, while some of the stars of the Basel5 cluster are also visible (its centre is on the left, outside of the image, see \citet{Kharchenko2013}). As indicated by the white cross, North is up and East is left.}
    \label{fig:ROMEfield}
\end{figure}

\begin{figure}[!]
  \centering
    %\subfloat{
       \includegraphics[width=0.9\textwidth]{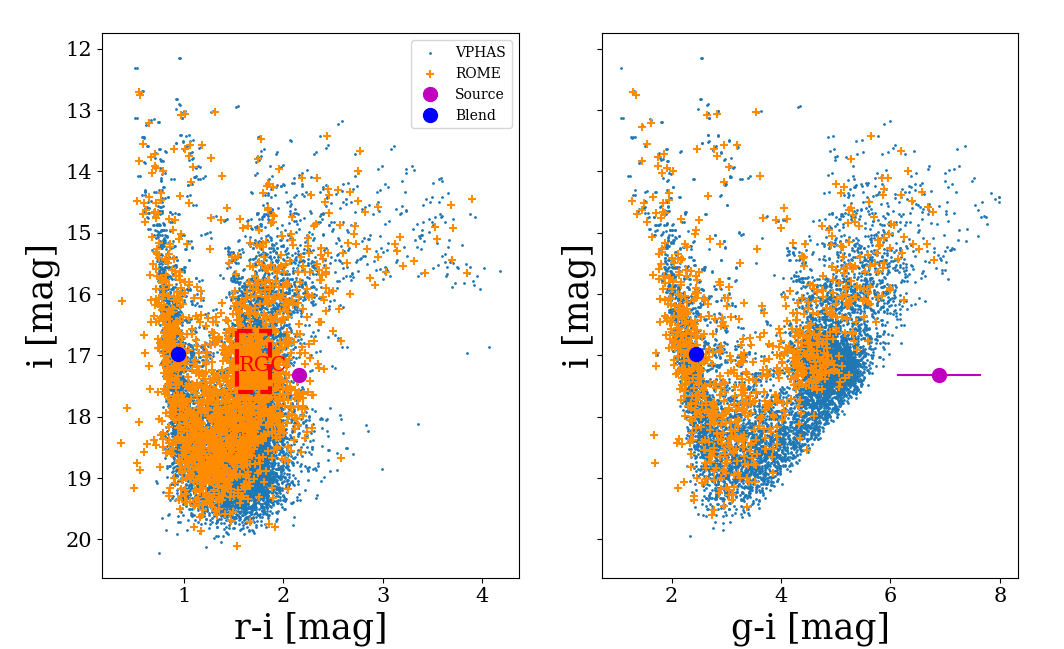}%}

  \caption{Color-Magnitude diagrams from the ROME/REA survey. The blue dots represents all stars within 2' around the event location from the VPHAS catalog \citep{VPHAS2016}, while the aligned ROME/REA stars are in orange. The source and blend are represented in magenta and blue, respectively. The dashed square on the left plot represents the stars that have been used to estimate the RGC centroid for both CMDs.}
\label{fig:ROMECMD}
\end{figure}

The ROME strategy consists of regular monitoring of 20 fields in the Galactic bulge in SDSS-g$^{\prime}$, SDSS-r$^{\prime}$ and SDSS-i$^{\prime}$, as described in \citet{Tsapras2019}, and is designed to improve our understanding of the source and blend properties. The photometry is obtained using the pyDANDIA algorithm (Street et al. in prep\footnote{https://github.com/pyDANDIA/pyDANDIA}) and calibrated to the VPHAS+ catalog \citep{VPHAS2016}. For this event, we investigated all combinations of filters and telescope sites and selected LSC\_A (i.e., LCO dome A in Chile) for the ROME CMD analysis, as it provided the deepest catalog. Figure~\ref{fig:ROMECMD} presents the CMD for stars in a 2'x2' square centered on the target, while Figure~\ref{fig:ROMEfield} presents a composite image of the ROME observations from LSC\_A. The latter presents the variable extinction in the field of view, as well as the two clusters NGC~6451 and Basel~5.

The first step is to estimate the centroid of the Red Giant Clump (RGC). In Figure~\ref{fig:ROMECMD}, stars located within 2' from the event location from the ROME/REA and VPHAS catalogs are displayed in (r-i,i) and (g-i,i) CMDs. We note that the location of the RGC is quite uncertain in the g-band for the ROME/REA data. This is due to the high extinction along this line of sight and leads to an accurate g-band calibration of the ROME data. We use the VPHAS magnitudes of these stars to estimate the centroid positions of the RGC in the three bands. We found the magnitudes of the RGC to be $((g),(r),i)_{RGC} = (21.989\pm0.007,18.836\pm0.005,17.139\pm0.005)$ mag.   

Following the method of \citet{Street2019} and using the intrinsic magnitude of the RGC $(M_g,M_r,M_i)_{o,c} = (1.331\pm0.056,0.552\pm0.026,0.262\pm0.032)$ \citep{Ruiz2018}, one can estimate the reddening $E(g-i)=3.74\pm0.1$ mag, $E(r-i)=1.41\pm0.1$ mag and extinction $A_i=2.3\pm0.1$ mag. The best model returns a source magnitude of $(g,r,i)_s = (24.2\pm0.8,19.47\pm0.05,17.31\pm0.03)$ mag. Because the event occured at the begining of the season, the event was poorly covered by the ROME data in the g-band and the source brightness is not well constrained. However, we can use the (r-i) color and i magnitude to estimate $\theta_*$. As described in Appendix~\ref{sec:boyajian_new}), we used the same catalog as \citet{Boyajian2014} to construct a new color-radius relation:
\begin{equation}
\log_{10}(2\theta_*) = (-0.298\pm0.044)(r-i)_o^2+(0.919\pm0.058)(r-i)_o-0.2i_o+(0.767\pm0.010)  
\end{equation}
This relation returns $\theta_*=9.8\pm1.2~\mu as $ while the second relation using the g-band returns $\theta_*=22.9\pm8.4~\mu as $. The latter value is inaccurate due to large uncertainties in the source brightness in the g-band. 

\subsection{MOA color-Magnitude Diagram}
\label{sec:moa_cmd}
\begin{figure}[h]
    \centering
    \includegraphics[width=0.9\columnwidth]{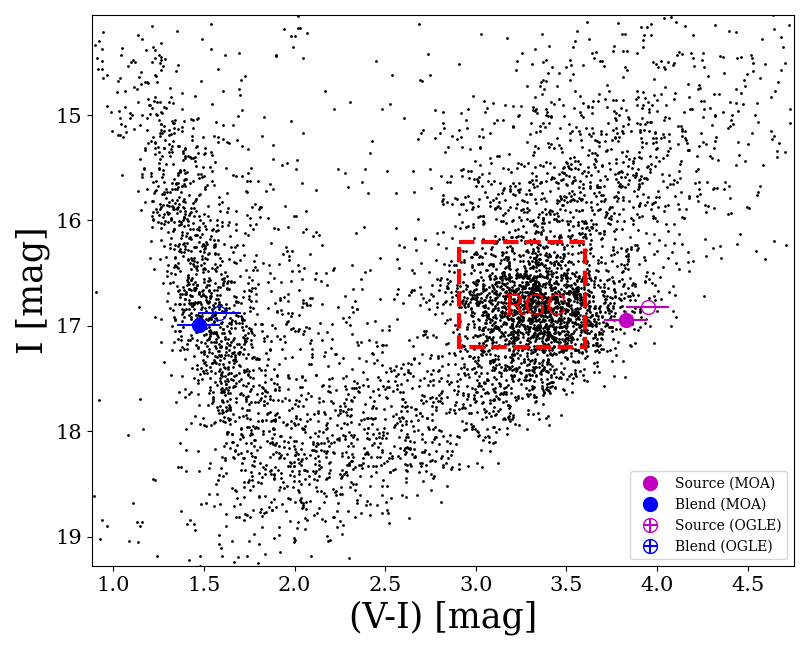}
    \caption{Color-Magnitude Diagram from the MOA survey, calibrated to the OGLE III system, for stars located in the square 2'x2' around the event. We also display the position of the source and the blend using the I measurements from the OGLE lightcurve.}
    \label{fig:CMD_MOA}
\end{figure}

The MOA magnitude system can be transformed to the OGLE-III magnitude system (i.e., the Johnson-Cousins system) by using the relation presented in Appendix~\ref{sec:mag_trans}. Using the intrinsic color and magnitude of the RGC $((V-I)_0,I_0) = (1.06,14.32)$ mag \citep{Nataf2013,Bensby2013} and subtracted to the measured position the RGC centroid in the Figure~\ref{fig:CMD_MOA}, we could estimate $(E(V-I),A_I)=(2.3\pm0.1,2.4\pm0.1)$ mag, in good agreement with the  previous estimation. Knowing that the transformed magnitudes of the source are $(V,I)_s = (20.8\pm0.1,16.94\pm0.08)$ mag, we found $(V,I)_{0,s} = (16.1\pm0.1,14.54\pm0.08)$ and we ultimately estimate $\theta_*=8.9\pm1.3~\mu as$ \citep{Adams2018}, in relative agreement with the previous estimation.
In Figure~\ref{fig:CMD_MOA}, we also display the source and blend position using the measurements from OGLE-IV in the I band and the transformed MOA V band. We estimate the source to be $(V,I)_s = ( (20.8\pm0.1,16.88\pm0.01)$ mag and derive $\theta_*=10.4\pm1.6~\mu as$. This estimate is likely more accurate than the previous, because it relies on a single color transformation (with the highest color term in Equation~\ref{eq:MOA_IOIII}).

\subsection{Gaia EDR3}
\label{sec:gaia_cmd}

\begin{figure}[h]
    \centering
    \includegraphics[width=0.9\columnwidth]{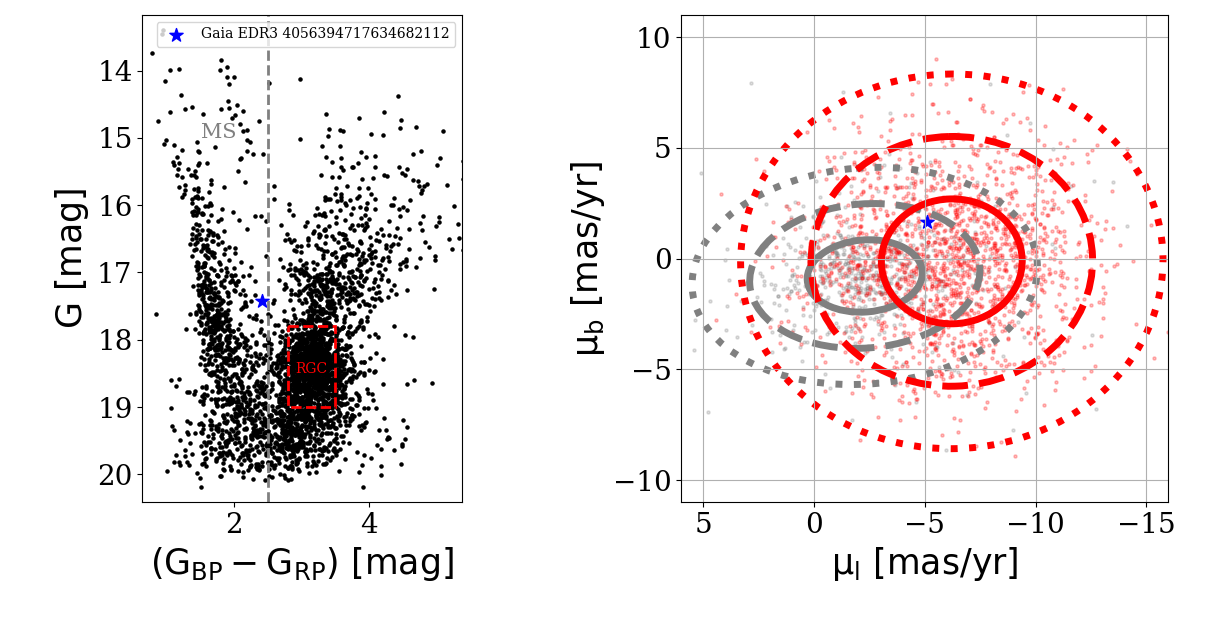}
    \caption{(Left) CMD of stars located 3' around the event, from the Gaia EDR3 release \citep{Brown2020}. The blue star is a multiple object containing the source, the lens and potentially additional blends}. (Right) Proper motion of these stars in galactic coordinates. The blue star indicates the position of the object located at the coordinates of the event, while red and grey points and ellipses indicate the Bulge and Disk population, respectively.
    \label{fig:gaiadr3}
\end{figure}

The Gaia mission \citep{Prusti2016} recently released their ``Early Data Release 3" data set (EDR3,\citep{Brown2020}), which significantly increases the volume and precision of the Gaia catalog. We queried the Gaia catalog requesting all stars
within a 3' radius\footnote{https://gaia.ari.uni-heidelberg.de/index.html} around the coordinates of the event to generate a Gaia CMD, which we present in Figure~\ref{fig:gaiadr3}. We limit our study to stars with a Re-normalised
Unit Weight Error (RUWE, a statistical criterion of the data quality) better than 1.4\footnote{https://gea.esac.esa.int/archive/documentation/GDR2/Gaia\_archive/chap\_datamodel/sec\_dm\_main\_tables/ssec\_dm\_ruwe.html}.
MOA-2019-BLG-008 is in the catalog (at 63 mas, Gaia EDR3 4056394717636682112) and appears in a sparse location of the CMD. The reported parallax is $p=0.39\pm0.12$ mas, which corresponds to a distance of $D=2.56\pm0.79$ kpc.
This object is also significantly redder and brighter than the blend discussed in the previous section. Indeed, by using the magnitude transformation from Gaia to the Johnson-Cousins
system\footnote{https://gea.esac.esa.int/archive/documentation/GDR2/} \citep{Bachelet2019}, we found this object to be $((V-I),I) = (2.25\pm0.07,16.17\pm0.05)$ mag,
and this is very likely the sum of the source and the blend previously discussed $((V-I),I)_{tot}=(2.30,16.19)$ mag. 
This is confirmed by several useful metrics available in the catalog. First, we compute the corrected BP and RP excess flux factor \citep{Evans2018,Riello2020} and find
0.28\footnote{https://github.com/agabrown/gaiaedr3-flux-excess-correction}, which corresponds to a blend probability of $\sim 0.3$ (see Figure 19 of \citet{Riello2020}. Secondly, we note that the fraction $R$ of visits that indicated a significant blend  (defined by `phot\_bp\_n\_blended\_transits' and `phot\_rp\_n\_blended\_transits' for the BP and RP bands respectively) divided by the number of visits used for the astrometric solution ('astrometric\_matched\_transits') is very high for both bands $R=36/39\sim 90\%$.

Following \citet{Mroz2020}, we plot in Figure~\ref{fig:gaiadr3} the distribution of galactic proper motions of the Disk (gray) and Bulge (red) populations. Note that Gaia EDR3 provides proper motion in equatorial coordinates which we transformed using the same method as in \citet{Bachelet2019}. The Disk population, approximated by the main sequence population, is estimated from the Gaia CMD by using all stars with $G_{BP}-G_{RP}\le2.5$. The Bulge population is estimated from the RGC population of the CMD (i.e., $2.8<G_{BP}-G_{RP}\le3.5$ and $17.8\le G\le 19$). Because this object is blended, it is difficult to extract meaningful constraints from the proper motion distribution. However, it will be possible to do so when the source and lens will be sufficiently separated, as discussed later. 

%just for reference
%\begin{tabular}{|l|r|r|r|r|r|r|r|r|r|r|r|l|r|r|r|r|r|r|r|}
%\hline
%  \multicolumn{1}{|c|}{source\_id} &
%  \multicolumn{1}{c|}{ref\_epoch} &
%  \multicolumn{1}{c|}{ra} &
%  \multicolumn{1}{c|}{ra\_error} &
%  \multicolumn{1}{c|}{dec} &
%  \multicolumn{1}{c|}{dec\_error} &
%  \multicolumn{1}{c|}{parallax} &
%  \multicolumn{1}{c|}{parallax\_error} &
%  \multicolumn{1}{c|}{pmra} &
%  \multicolumn{1}{c|}{pmra\_error} &
%  \multicolumn{1}{c|}{pmdec} &
%  \multicolumn{1}{c|}{pmdec\_error} &
%  \multicolumn{1}{c|}{duplicated\_source} &
%  \multicolumn{1}{c|}{phot\_g\_mean\_flux} &
%  \multicolumn{1}{c|}{phot\_g\_mean\_flux\_error} &
%  \multicolumn{1}{c|}{phot\_g\_mean\_mag} &
%  \multicolumn{1}{c|}{phot\_bp\_mean\_mag} &
%  \multicolumn{1}{c|}{phot\_rp\_mean\_mag} &
%  \multicolumn{1}{c|}{bp\_rp} &
%  \multicolumn{1}{c|}{bp\_g}  \\
%\hline
%  4056394717634682112 & 2015.5 & 267.9829897253411 & 0.15440841726026935 & -29.989764233927147 & 0.12432716993142232 & 0.6462484902692556 & 0.15958954987787757 & -3.9945708996936347 & 0.2832867530255288 & -3.302831560381608 & 0.2206411686230916 & false & 2001.3995974610484 & 3.4647626808923073 & 17.435032 & 18.330833 & 15.883127 & 2.4477062 & 0.89580154  \\
%\hline\end{tabular}
\subsection{Analysis of the blend}
\label{sec:blend}
As can be seen in the different CMD's, there is a significant blend flux in the datasets, which likely belongs to the population of foreground stars of the Galactic Disk. The measurements from MOA/OGLE are $(V,I)_{b} = (18.5\pm0.1,16.88\pm0.01)$ mag and are consistent with a late F dwarf located at $\sim$ 2.5 kpc \citep{Bessell1988,Pecault2013} (assuming half extinction). Measurements from the ROME survey indicate that the blend brightnesses are $(g,r,i)_{b} = (19.43\pm0.01,17.91\pm0.01,16.97\pm0.02)$ mag, consistent with a G dwarf at $\sim 2.0$ kpc \citep{Finlator2000,Schlafly2010}. These results are in agreement with the lens properties derived in the next section.

The source and blend have similar brightness in Cousins $I$ band, but the former is much redder. Therefore, the object identified at this location by Gaia is dominated by light from the blend. At the epoch J2016, Gaia measures a total offset of $60\pm15$ mas with respect to the event location measured in 2019 (i.e., during peak magnification). The reported error on the distance has been computed from the North and East components error from ground surveys (of the order of $\sim$ 15 mas) and neglecting Gaia errors (of the order of $\sim$ 0.1 mas). Similarly, the magnified source and the baseline object in the KMTNet images are separated by $\sim$0.068 pixels, which is equivalent to 27 mas. Because the blend and source have similar brightness, this indicates a separation between the blend and the source of $\sim$ 60 mas. Therefore, the hypothesis that the blend, or a potential companion, is the lens is probable. Assuming that the blend is the lens, $\piE\sim0.2$, a source distance $D_s\sim8$ kpc and that the light detected by Gaia is solely due to the blend, we can estimate $\thetaE\sim(0.39-0.125)/0.2\sim1.3$ mas and $M_l\sim0.8 \rm{M_\odot}$. So the astrometric solution is also compatible with a G dwarf at $\sim$ 2.5 kpc, with the notable exception of the relative proper motion. Indeed, we can estimate the geocentric relative proper motion to be $\mu_{geo}=\theta_E/t_E\sim1.3/80\sim6$ mas/yr. Because this event peaked in early March, the heliocentric correction $v_{\oplus}\pi_{rel}/\rm{au}\sim(0.2,-0.4)$ is small and we can therefore assume $\mu_{geo}=\mu_{hel}$ \citep{Dong2009}. The separation between the lens and the source at the Gaia epoch (J2016) is therefore expected to be $\sim$ 18 mas, significantly smaller than the previously measured $\sim$60 mas. However, this argument can not, by itself, rule out the hypothesis that the blend is the lens due to the relatively large errors.

Therefore, both photometric and astrometric arguments provide sufficient evidence that the lens represents a significant fraction of the blended light, but only high-resolution imaging in the near future will provide a conclusive answer to this puzzle.

\begin{table}[h!]

\centering
\caption{Source and blend magnitudes for the three 2L1S models (results are almost identical for $u_0<0$ and $u_0>0$). Numbers in brackets represent the 1$\sigma$ errors. MOA magnitudes have been converted to the OGLE-III system using the transformation in the Appendix~\ref{sec:mag_trans}.} \label{tab:source_blend}
\begin{tabular}{cccccccc}
\hline
\hline
&Filter&\multicolumn{2}{c}{Static}&\multicolumn{2}{c}{Parallax}&\multicolumn{2}{c}{Parallax+Orbital Motion}\\
&&Source&Blend&Source&Blend&Source&Blend \\
\hline
ROME & g$^{\prime}$ & {\bf 24.2(0.8)} & {\bf 19.43(0.01)} &{\bf 24.2 (0.8)}&{\bf 19.43(0.01)}&{\bf 24.0(0.8)}&{\bf 19.43(0.01)}\\
ROME & r$^{\prime}$ & {\bf 19.49 (0.05)} & {\bf 17.91(0.01)} &{\bf 19.47(0.05)}&{\bf 17.92(0.01)}&{\bf 19.27(0.05)}&{\bf 17.97 (0.02)}\\
ROME & i$^{\prime}$ & {\bf 17.34 (0.03)} & {\bf 16.97(0.02)} &{\bf 17.31(0.03)}&{\bf 16.98(0.02)}&{\bf 17.11 (0.03)}&{\bf17.15 (0.03)}\\
\hline
OGLE & I & 16.857 (0.002) & 16.844(0.002) &16.824(0.002)&16.876(0.002)&16.642(0.002)&17.108(0.003)\\
\hline
MOA & I & 17.0 (0.1) & 17.0(0.1) &16.9(0.1)&17.0(0.1)&16.8(0.1)&17.2(0.1)\\
MOA & V & 20.8 (0.1) & 18.5(0.1) &20.8(0.1)&18.5(0.1)&20.6(0.1)&18.5(0.1)\\
\hline

\end{tabular}
\end{table}

\section{Galactic models}
\label{sec:galmod}

Because the normalised radius of the source $\rho$ can not be estimated from the fit, inferring the Einstein radius is not possible without extra measurements, such as the microlensing astrometric signal \citep{Dominik2000}, the lens flux or the lens and source separation measurements after several years \citep{Alcock2001, Beaulieu2019}. In order to estimate the physical properties of the lens, prior information from Galactic models can be used. By drawing random source-lens pairs from distributions of stellar physical parameters derived from the galactic models along the line of sight, and calculating the respective microlensing model parameters, the lens mass and distance probability densities can be estimated. This has been done many times in the past with parameterized models specifically designed to study microlensing events \citep{Han1995,Han2003,Dominik2006,Bennett2014,Koshimoto2021}. But there are also modern galactic models that have been extensively tested and are publicly accessible. From a theoretical point of view, these elaborate models are of great interest because they are including more relevant quantities such as color, extinction and stellar type, for instance. These quantities can be used to constrain physical parameters, but also to predict properties for follow-up observations in the more distant future. In this work, we performed a parallel analysis using the parametric model of \citet{Dominik2006}, the Besan\c{c}on model and the GalMod model described thereafter.  %In this work, instead of using the parameterized models that are commonly used in the literature \citep{Han1995,Han2003,Dominik2006,Bennett2014}, we compare the results from two public Galactic models.

\subsection{The Besan\c{c}on Model}
The first galactic model we use to generate a stellar population is the Besan\c{c}on Model\footnote{https://model.obs-besancon.fr/modele\_descrip.php} \citep{Robin2003}, version M1612. This version consists of an ellipsoidal Bulge titled by $\sim 10^\circ$ from the Sun-Galactic center direction, and populated with stellar masses drawn from a broken power law initial mass-function (IMF) $dN/dM\propto M^\alpha$, with $\alpha=-1$ and $\alpha=-2.35$ for $0.15\le 0.7 M_\odot$ and $M>0.7 M_\odot$ respectively \citep{Robin2012,Penny2019}. The Disk is modelled by a thin disk component with a two-slope power law IMF, with $\alpha=1.6$ and $\alpha=3.0$ for $M\le1 M_\odot$ and $M>1 M_\odot$\citep{Robin2012}, while the density distribution is derived from \citet{Einasto1979}. The outer part of the disk model has recently been updated and is described in \citet{Amores2017}. The thick disk and halo population are fully described in \citet{Robin2014}, while the kinematics of the population are described in \citep {Bienayme2015}. We select the \citet{Marshall2006} 3D map to estimate the extinction for the simulation. Finally, we note that the Besan\c{c}on Model has been used in several studies for microlensing predictions. Based on the original work of \citet{Kerins2009}, \citet{Awiphan2016} and \citet{Specht2020} developed the $\rm{MaB\mu ls-2}$ software that computes theoretical maps of the distribution of optical depth, event rate and timescales of microlensing events, that are in good agreement with observations. In particular, $\rm{MaB\mu ls-2}$ predictions of event rate and optical depth are excellent agreement with the 8 years of observations from the OGLE survey \citep{Mroz2019}. The Besan\c{c}on Model has also been used by \citet{Penny2013} to simulate the potential yields of a microlensing exoplanet survey with the Euclid space telescope. More recently, \citet{Penny2019} and \citet{Samson2020} used an updated version of the Besan\c{c}on Galactic model to estimate the expected number of detections of bound and unbound planets from the Roman (formerly known as WFIRST) microlensing survey \citep{Spergel2015}.

\subsection{GalMod}

The second simulation was made using the “Galaxy Model” (GalMod, version 18.21), which is a theoretical stellar population synthesis model \citep{pasetto2018a}\footnote{\url{https://www.galmod.org/gal/}} simulating a mock catalog for a given field of view and photometric system. Similarly as for the Besan\c{c}on model, the parameter range in magnitude and color permits the simulation of faint lens stars down to the dwarf and brown dwarf regime. Briefly, GalMod consists of the sum of several stellar populations including a thin and a thick disk, a stellar halo, and a bulge immersed in a halo of dark matter. Stars are generated using the multiple-stellar population consistency theorem described in \citet{Pasetto2019} with a kinematics model from \citet{Pasetto2016}. For our simulation, we used the Rosin-Rammler star formation rate (SFR) \citep{Chiosi1980} for the Bulge and the tilted bar. The thin disk is a combination of five different stellar populations with various ages and kinematics, with a constant SFR, while the thick disk is drawn from a single population. We used the same IMF for all the different components of the model \citep{Kroupa2001}.

\subsection{Methodology}
We first requested samples from the two models within a $2\prime \time 2\prime$ cone along the line of sight to the event, and set the maximum distance to 10 kpc. 
We then draw samples of lens and source star combinations and apply a sequence of rules. First, the source has to be more distant than the lens. Then, we consider an event only if the angular separation between the source and the lens is below 10". Following the approach described in \citet{Shin2019}, we proceed to compute the associated event parameters (i.e., $t_{\rm{E}}$, $\piE$, $\theta_*$ and $I_s$ in this case) and compare them with our measured observables derived from modeling. Each such combination contributes to the final derived distribution with a weight $w_i=\frac{1}{\sqrt{2\pi|C^{-1}|}}\exp(-\delta_i^2/2)$, with $\delta_i^2$ being the Mahalanobis distance:
\begin{equation}
    \delta_i^2 = \mathbf{r_i}^\intercal\mathbf{C^{-1}}\mathbf{r_i}
\end{equation}
where $\mathbf{r_i}=(t_E-t_{E,i},\piE-\pi_{E,i},\theta_*-\theta_{*,i},I_s-I_{s,i})$ are the differences between the best fit model parameters and the simulated parameters and $\mathbf{C}$ is the covariance matrix. Note that we also reject models with $\rho\ge0.01$, following the discussion presented in the Section~\ref{sec:norho}.

Because the galactic models return a finite number of stars (168134 for the Besan\c{c}on model and 64679 for the GalMod model) and the event parameters are slightly unusual (with $t_E\sim80$ days and $\theta_*\sim10~\mu as$), a large fraction of lens and source combination have a null weight. For instance, the Besan\c{c}on model contains only $\sim0.4$ \% of stars with $|\theta_*-11.5|<3\sigma$ and about 1\% of events are expected to have $t_E\sim80$ days \citep{Mroz2019}. Therefore, the vast majority of trials ($\ge 99.9$ \%) have null weights $w_i$ and it would therefore require several thousands of billions of trials to obtain meaningful parameter distributions. 
In light of this, we adjusted our strategy and adopted an MCMC approach. Using the Mahalanobis as the log-likelihood, we adapt the galactic models to define priors on the modeling parameters: the source and lens distances $D_s$ and $D_l$, the proper motion of the source and lens, the mass of the lens $M_l$, the angular radius of the source $\theta_*$ and the magnitude of the source $I_s$. We use a Kernel Density Estimation (KDE) algorithm to derive continuous distributions from the galactic-model samples. This allows a prior estimation across the entire parameter space, but at the cost of a somewhat smoother distribution and the use of extrapolation. 

\subsection{Results}
\begin{figure}[h]
    \centering
    \includegraphics[width=0.9\columnwidth]{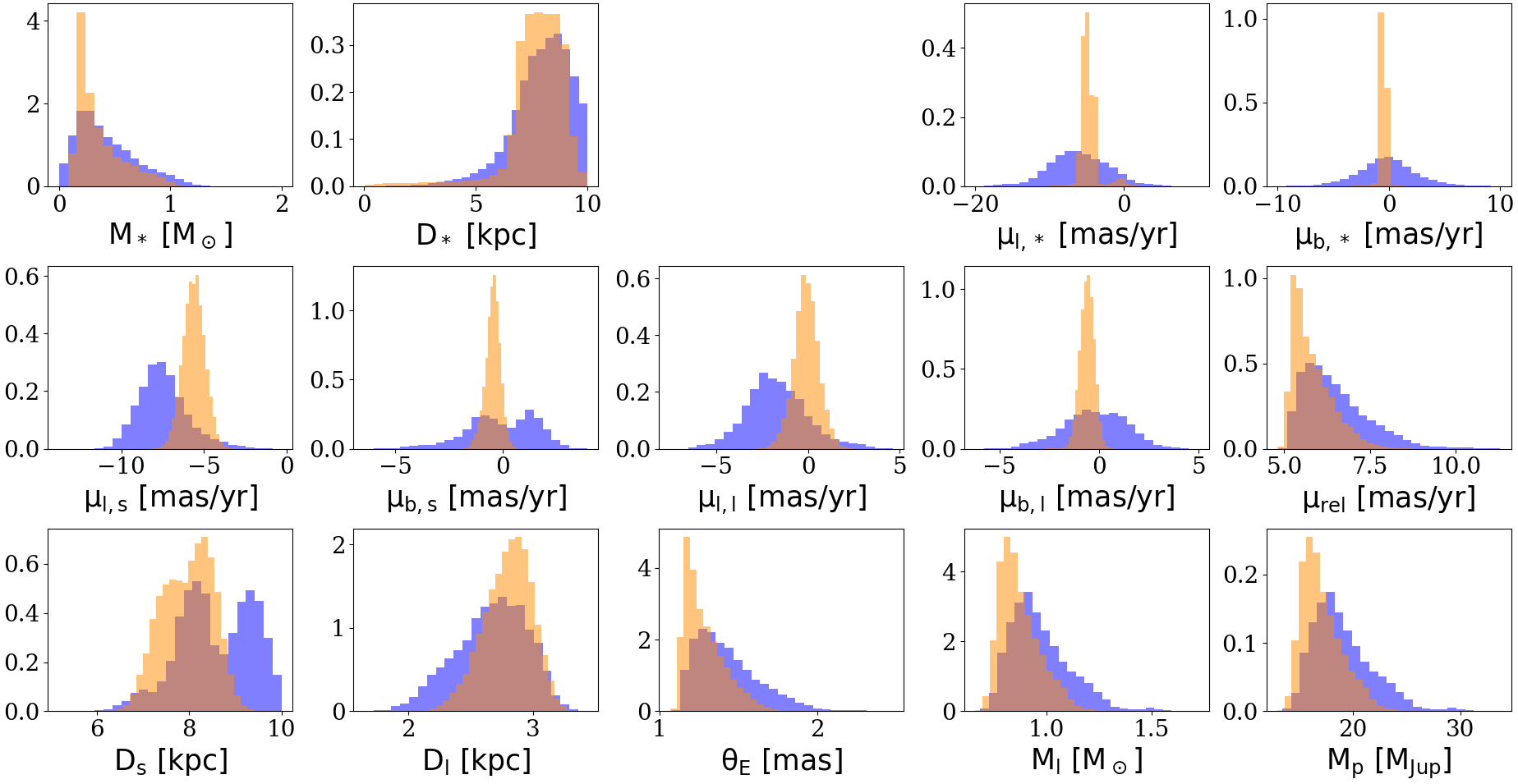}
    \caption{(Normalized) Distribution of the derived parameters for the 2L1S+POM model. The first line represents the mass, distance and proper motion distributions of stars from the Besan\c{c}on (blue) and GalMod (orange) models. The last two lines represent the posterior distribution of the event parameters from the MCMC exploration.}
    \label{fig:galactics}
\end{figure}

We present the posterior distribution for the model 2L1S+POM in Figure~\ref{fig:galactics} and the derived results for all models in Table~\ref{tab:distresults}. Results from the Galactic model of \citet{Dominik2006} is also presented for comparison. As a supplementary control, we also derived posterior distributions from the recent Galactic model of \citet{Koshimoto2021}, especially designed for microlensing studies, and found consistent results.

Despite the relatively broad distributions, we find that galactic models are in good agreement for all models. The main differences are seen in the distribution of the source and lens proper motions. The GalMod model has a much narrower distribution of stellar proper motions, as can be seen in the first line of Figure~\ref{fig:galactics}, which directly propagates to the source and lens proper motions. However, the relative proper motion of the two galactic models are in 1-$\sigma$ agreement, at $\sim$5.5 mas/yr. This is because the relative proper motion is strongly constrained by $t_{\rm{E}}$, which is well determined from the models in this case. For all microlensing models, the derived mass and distance of the host is compatible with the measured blend light, with the exception of the 2L1S-P model, which is much fainter with $V_l\sim20.5$. The companion is an object at the planet/brown dwarf boundary.

While the binary source and static binary models can be safely discarded due to their very high $\Delta\chi^2$ values relative to the best-fit model, the selection of the best overall model with/without orbital motion ($\Delta\chi^2\sim200$) and the sign of $u_0$ ($\Delta\chi^2\sim50$) is less trivial. In principle, the $\Delta\chi^2$ between the various models is statistically significant. However, despite error-bar rescaling, data set residuals can be affected by low level systematics, leading to errors that are not normally distributed \citep{Bachelet2017}. Because they modify the source trajectory in a similar way, the orbital motion and parallax parameters are often correlated \citep{Batista2011} which is also the case for this event. The North component $\pi_\mathrm{E,N}$ of the parallax-vector is in agreement between all non-static models, suggesting that the parallax signal is strong and real, which is expected for an event duration of $t_E\sim80$ days. For models including orbital motion, we can use the results from galactic models to verify if the system is bound or not \citep{Dong2009,Udalski2018}. The condition for bound systems is $(K_E+P_E)<0$, where $K_E$ and $P_E$ are the kinetic and potential energies, and this can be rewritten in terms of (projected) escape velocities ratio \citep{Dong2009}:
\begin{equation}
{{v_\perp^2}\over{v_{esc,\perp}^2}} = {{(a_\perp)^3}\over{8\pi^2M_l}}\biggl[\biggl({{1}\over{s}}{{ds}\over{dt}}\biggr)^2+\biggl({{d\alpha}\over{dt}}\biggr)^2\biggr]\le1
\end{equation}

 We found ${{v_\perp^2}\over{v_{esc,\perp}^2}} \sim 4$ and ${{v_\perp^2}\over{v_{esc,\perp}^2}} \sim 6$ for the 2L1S-POM and 2L1S+POM models, respectively. Taken at face values, the ratio of projected velocities indicates that the companion is not bound and that these models are unlikely. However, the relative errors are large, i.e. $ \ge100\%$, and therefore the models with orbital motion can not be completely ruled out. But given the relatively low improvement in the $\chi^2$ of the orbital motion models, we decided to not explore more sophisticated models, such as the full Keplerian parametrization \citep{Skowron2011}.

\begin{table}[h!]
\scriptsize
\caption{Derived parameters from the galactic models and the event modeling, defined as the 16, 50 and 84 MCMC chains percentile. Columns display the event physical parameters while rows correspond to the different models with parallax and orbital motion presented in Section~\ref{sec:modeling}. For each microlensing model, the first rows are results from the Besan\c{c}on model, the second rows are from the GalMod model and the third rows represent results from \citet{Dominik2006}.}  
\label{tab:distresults}
\begin{tabular}{lcccccccccc}
\hline
\hline
Models &$D_s ~\rm{[kpc]}\footnote{The source distance and errors have been fixed for \citet{Dominik2006}}$&$\mu_{l,s}$ [mas/yr] & $\mu_{b,s}$ [mas/yr] &
$D_l ~\rm{[kpc]}$ &$\mu_{l,l}$ [mas/yr] &$\mu_{b,l}$ [mas/yr] &$\mu_{rel}$ [mas/yr] &$\thetaE$ [mas] &$M_l ~\rm{[M_\odot]}$ &$M_p ~\rm{[M_{Jup}]}$ \\
\hline
 &$6.7^{+1.8}_{-1.1}$&$-4.3^{+1.2}_{-2.0}$& $-1.1^{+1.5}_{-2.4}$ &
$2.2^{+0.2}_{-0.3}$&$0.6^{+1.4}_{-2.1}$&$-0.5^{+1.1}_{-2.5}$&$5.4^{+1.1}_{-0.6}$& $1.2^{+0.2}_{-0.1}$ &$0.6^{+0.1}_{-0.1}$&$25^{+5}_{-3}$\\

2L1S-P &$8.0^{+0.6}_{-0.6}$&$-5.2^{+0.7}_{-0.8}$& $-0.4^{+0.4}_{-0.4}$ &
$2.3^{+0.2}_{-0.2}$&$-0.2^{+0.6}_{-0.7}$&$-0.6^{+0.5}_{-0.5}$&$5.2^{+0.8}_{-0.4}$& $1.2^{+0.2}_{-0.1}$ &$0.6^{+0.1}_{-0.1}$&$24^{+4}_{-2}$\\

 &$8.0\pm0.5$&*&*&
$2.6^{+1.2}_{-0.9}$&*&*&*&*&$0.5^{+0.4}_{-0.2}$&$21^{+15}_{-6}$\\
\hline
 &$6.6^{+1.3}_{-0.9}$&$-3.9^{+1.4}_{-2.2}$& $-0.8^{+1.1}_{-1.9}$ &
$3.2^{+0.3}_{-0.4}$&$0.4^{+1.4}_{-1.8}$&$-1.2^{+2.1}_{-1.5}$&$4.8^{+0.8}_{-0.4}$& $1.2^{+0.2}_{-0.1}$ &$1.1^{+0.2}_{-0.1}$&$36^{+6}_{-4}$\\

2L1S+P &$8.0^{+0.6}_{-0.6}$&$-5.2^{+0.7}_{-0.7}$& $-0.4^{+0.4}_{-0.4}$ &
$3.5^{+0.3}_{-0.4}$&$-0.3^{+0.8}_{-0.6}$&$-0.3^{+0.8}_{-0.6}$&$5.0^{+1.1}_{-0.4}$& $1.2^{+0.3}_{-0.1}$ &$1.1^{+0.3}_{-0.1}$&$37^{+10}_{-4}$\\

 &$8.0\pm0.5$&*&*&
$2.6^{+1.2}_{-0.9}$&*&*&*&*&$0.8^{+0.4}_{-0.4}$&$27^{+12}_{-13}$\\
\hline
 &$8.3^{+1.0}_{-0.9}$&$-7.5^{+1.7}_{-1.4}$& $0.0^{+1.6}_{-1.9}$ &
$2.4^{+0.3}_{-0.3}$&$-0.3^{+0.8}_{-0.6}$&$-1.4^{+1.5}_{-1.7}$&$6.3^{+1.4}_{-0.7}$& $1.4^{+0.3}_{-0.2}$ &$0.8^{+0.2}_{-0.1}$&$17^{+4}_{-2}$\\

2L1S-POM &$8.0^{+0.5}_{-0.7}$&$-5.6^{+0.6}_{-0.7}$& $-0.4^{+0.3}_{-0.4}$ &
$2.6^{+0.2}_{-0.2}$&$0.0^{+0.5}_{-0.6}$&$-0.1^{+0.6}_{-0.6}$&$5.7^{+0.6}_{-0.3}$& $1.2^{+0.1}_{-0.1}$ &$0.7^{+0.1}_{-0.1}$&$15^{+2}_{-1}$\\

&$8.0\pm0.5$&*&*&
$2.6^{+1.2}_{-0.9}$&*&*&*&*&$0.6^{+0.3}_{-0.3}$&$12^{+6}_{-6}$\\
\hline
 &$8.5^{+1.0}_{-0.7}$&$-7.6^{+1.5}_{-1.3}$& $0.0^{+1.6}_{-1.5}$ &
$2.7^{+0.3}_{-0.3}$&$-1.8^{+1.7}_{-1.4}$&$-0.2^{+1.6}_{-1.6}$&$6.3^{+1.1}_{-0.7}$& $1.4^{+0.3}_{-0.2}$ &$0.9^{+0.2}_{-0.1}$&$18^{+3}_{-2}$\\

2L1S+POM &$8.0^{+0.5}_{-0.6}$&$-5.6^{+0.6}_{-0.7}$& $-0.4^{+0.3}_{-0.4}$ &
$2.8^{+0.2}_{-0.2}$&$-0.1^{+0.7}_{-0.6}$&$-0.6^{+0.3}_{-0.4}$&$5.6^{+0.7}_{-0.4}$& $1.3^{+0.2}_{-0.1}$ &$0.9^{+0.1}_{-0.1}$&$17^{+2}_{-1}$\\

 &$8.0\pm0.5$&*&*&
$2.6^{+1.2}_{-0.9}$&*&*&*&*&$0.6^{+0.4}_{-0.3}$&$12^{+9}_{-5}$\\
\hline
\hline
\end{tabular}
\end{table}

\section{Conclusions}
\label{sec:concl}
We presented the analysis of the microlensing event MOA-2019-BLG-008. The modeling of this event supports a binary-lens interpretation with a mass ratio $q\lesssim 0.04$ between the two components of the lens. Because the source trajectory did not approach the caustics of the system, finite-source effects were not detected, so the lens mass and distance could only be weakly constrained. We used the Besan\c{c}on and GalMod synthetic stellar population models of the Milky Way to estimate the most likely physical parameters of the lens. By using samples generated by these models, in combination with available constraints on the event timescale $t_{\rm{E}}$, the microlensing parallax $\piE$, the source magnitude $I_s$ and angular radius $\theta_*$, we were able to place constraints on the lens mass and distance. We found that all galactic models, including the one from \citet{Dominik2006}, converge to similar solutions for the lens mass and distance, despite different hypotheses (especially for stellar proper motions). We explore several microlensing binary lens models and they are all consistent with a main sequence star lens located at $\le$ 4 kpc from Earth. The microlensing models also indicate the presence of a bright blend, separated by $\Delta\sim60$ mas from the source, with $\sim (V,I)_b=(18.5,17.0)$ mag. Assuming that the blend suffers half of the total extinction towards the source, this object is compatible with a late F-dwarf at $\sim 2.5$ kpc \citep{Bessell1988,Pecault2013}, consistent with the lens properties derived from the galactic models analysis. The astrometric measurement made by Gaia at this position returns $D=2.56\pm0.79$ kpc. Assuming this object to be the lens, we derived $\theta_E\sim1.3$ mas and $0.8 M_\odot$, also consistent with the previous estimations. Depending on the exact nature of the host, the lens companion is either a massive Jupiter or a low mass brown dwarf. Given their relative proper motion, $\mu_{rel}=5.5$ mas/yr, the lens and source should be sufficiently separated to be observed via high-resolution imaging in about 10 years with 10-m class telescopes. This would provide the necessary additional information needed to confirm the exact nature of the lens, including the companion.

 Even though the physical nature of the host star cannot yet be firmly established, it is almost certain that the companion is located at the brown dwarf/planet mass boundary. The increasing number of reported discoveries of such objects, especially by microlensing surveys (see for example \citet{Bachelet2019} and references therein), provides important observational data which can be used to improve the theoretical framework underpinning planet formation. Indeed, there is more and more evidence that the critical mass to ignite deuterium (i.e., $\sim13~M_{Jup})$ does not represent a clear-cut limit \citep{Chabrier2014}. While there is compelling evidence that the two classes of objects are produced by different formation processes \citep{Reggiani2016,Bowler2020}, more observational constraints will be necessary in order to better appreciate the differences between them.

 Similarly to MOA-2019-BLG-008, it can be expected that a fraction of events detected by the Roman microlensing survey will miss at least one mass-distance relation, i.e. $\thetaE$ or $\piE$. In this context, the Besan\c{c}on and GalMod models can be particularly helpful in estimating the most likely parameters for the lens. Indeed, while \citet{Penny2019} and \citet{Terry2020} report some discrepancies between observations and their catalogs, these models are constantly upgraded to refine their predictions. In particular, the high accuracy astrometric measurements from Gaia will offer unique constraints on the proper motions and distances of stars up to the Galactic Bulge population at $\sim 8$ kpc. 
%% The reference list follows the main body and any appendices.
%% Use LaTeX's thebibliography environment to mark up your reference list.
%% Note \begin{thebibliography} is followed by an empty set of
%% curly braces.  If you forget this, LaTeX will generate the error
%% "Perhaps a missing \item?".
%%
%% thebibliography produces citations in the text using \bibitem-\cite
%% cross-referencing. Each reference is preceded by a
%% \bibitem command that defines in curly braces the KEY that corresponds
%% to the KEY in the \cite commands (see the first section above).
%% Make sure that you provide a unique KEY for every \bibitem or else the
%% paper will not LaTeX. The square brackets should contain
%% the citation text that LaTeX will insert in
%% place of the \cite commands.

%% We have used macros to produce journal name abbreviations.
%% \aastex provides a number of these for the more frequently-cited journals.
%% See the Author Guide for a list of them.

%% Note that the style of the \bibitem labels (in []) is slightly
%% different from previous examples.  The natbib system solves a host
%% of citation expression problems, but it is necessary to clearly
%% delimit the year from the author name used in the citation.
%% See the natbib documentation for more details and options.
\textit{Software: Astropy \citep{Astropy2018}, emcee \citep{Foreman2013},  pyLIMA \citep{Bachelet2017}}

\section*{Acknowledgements}
RAS and EB gratefully acknowledge support from NASA grant 80NSSC19K0291.  YT and JW acknowledge the support of DFG priority program SPP 1992 “Exploring the Diversity of Extrasolar Planets” (WA 1047/11-1).  KH acknowledges support from STFC grant ST/R000824/1. J.C.Y. acknowledges support from N.S.F Grant No. AST-2108414. Work by C.H. was supported by the grants of National Research
Foundation of Korea (2019R1A2C2085965 and 2020R1A4A2002885). This research has made use of NASA's Astrophysics Data System, and the NASA Exoplanet Archive.  The work was partly based on data products from observations made with ESO Telescopes at the La Silla Paranal Observatory under programme ID 177.D-3023, as part of the VST Photometric H{alpha} Survey of the Southern Galactic Plane and Bulge (VPHAS+, www.vphas.eu).  This work also made use of data from the European Space Agency (ESA) mission {\it Gaia} (\url{https://www.cosmos.esa.int/gaia}), processed by the {\it Gaia} Data Processing and Analysis Consortium (DPAC, \url{https://www.cosmos.esa.int/web/gaia/dpac/consortium}). Funding for the DPAC
has been provided by national institutions, in particular the institutions
participating in the {\it Gaia} Multilateral Agreement.  CITEUC is funded by National Funds through FCT - Foundation for Science and Technology (project: UID/Multi/00611/2013) and FEDER - European Regional Development Fund through COMPETE 2020 – Operational Programme Competitiveness and Internationalization (project: POCI-01-0145-FEDER-006922).   DMB acknowledges the support of the NYU Abu Dhabi Research Enhancement Fund under grant RE124.  This research uses data obtained through the Telescope Access Program (TAP), which has been funded by the National Astronomical Observatories of China, the Chinese Academy of Sciences, and the Special Fund for Astronomy from the Ministry of Finance. This work was partly supported by the National Science Foundation of China (Grant No. 11333003, 11390372 and 11761131004 to SM). This research has made use of the KMTNet system operated by the Korea
Astronomy and Space Science Institute (KASI) and the data were
obtained at three host sites of CTIO in Chile, SAAO in South Africa,
and SSO in Australia.The MOA project is supported by JSPS KAKENHI Grant Number JSPS24253004, JSPS26247023, JSPS23340064, JSPS15H00781, JP16H06287, and JP17H02871.

\appendix
\section{New SDSS color-radius relation}
\label{sec:boyajian_new}

\begin{figure}[h]
    \centering
    \includegraphics[width=0.9\columnwidth]{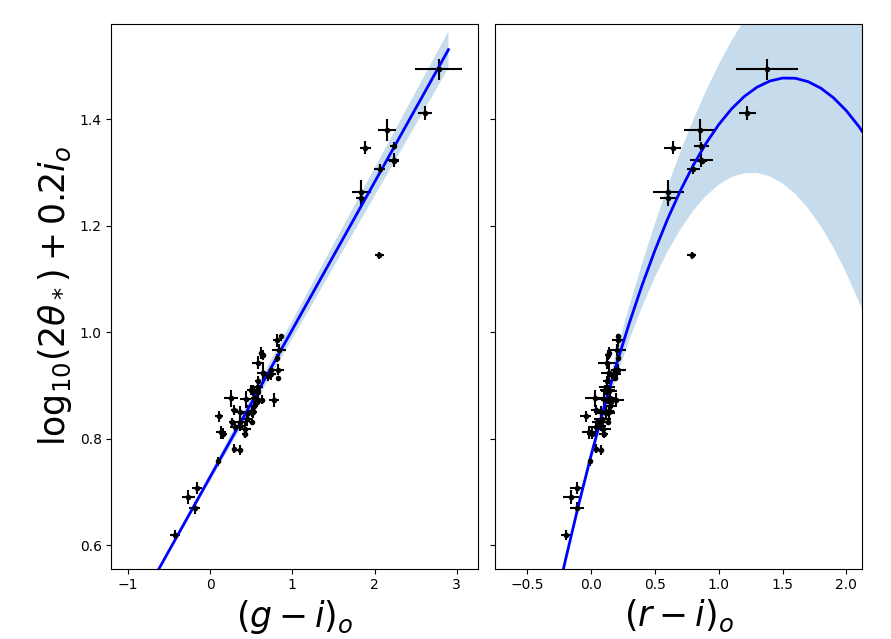}
    \caption{The new color-radius relation based on the $(g-i)_o$ (left) and $(r-i)_o$ (right) colors. For both figures, the best fit is indicated by the blue line and the 1 $\sigma$ uncertainties interval is displayed in light blue.}
    \label{fig:boyajian_new}
\end{figure}

As discussed in the main text, the source magnitude in the g-band from the ROME survey is not well known, due to the low sampling of the lightcurve. But the source brightness in the r and i bands are well measured. Because \citet{Boyajian2014} does not provide a relation for these bands, we decided to collect the data and estimate these relations. As described in \citet{Boyajian2014} and avalaible on Simbad\footnote{\url{http://simbad.u-strasbg.fr/simbad/sim-ref?querymethod=bib&simbo=on&submit=submit+bibcode&bibcode=2012ApJ...746..101B}}, we used the magnitudes measruements from \citet{Boyajian2013} and the angular diameter measurements from various interferometers: the CHARA Array \citep{DiFolco2004,Bigot2006,Baines2008,Baines2012,Boyajian2012a,Ligi2012,Bigot2011,vonBraun2011,Crepp2012,Bazot2011,Huber2012}, the Palomar Testbed Interferometer \citep{vanBelle2009}, the Very Large Telescope Interferometer \citep{Kervella2003a,Kervella2003b,Kervella2004c,DiFolco2004,Thevenin2005,Chiavassa2012}, the Sydney university Stellar Interformeter \citep{Davis2011}, the Narrabri Intensity Interferometer \citep{Hanbury1974}, Mark III \citep{Mozurkewich2003} and the Navy Prototype Oprical Interfereometer \citep{Nordgren1999,Nordgren2001}. Then, we fitted the color-radius relation as:
\begin{equation}
\log_{10}{2\theta_*} = \sum a_i(X^i)-0.2i_o    
\end{equation}
where $X_i$ is the considered color and $i_0$ is the de-reddened magnitude in the i-band. 
For stars that display several brightness measurements in the g, r and i bands, we used the mean as our final values, and the error was estimated from the sample variance with the (quadratic) addition of a 0.005 mag minimum error. We also used the error on the measured radius if available, and added quadratically the error on the observed $i_o$ magnitude. We explored several solutions using polynomials of different degrees and stopped as soon as the relative error on $a_i$ reached 1. Ultimately, we obtained:

\begin{equation}
\log_{10}(2\theta_*) = (-0.298\pm0.044)(r-i)_o^2+(0.919\pm0.058)(r-i)_o-0.2i_o+(0.767\pm0.010)  
\end{equation}

and 

\begin{equation}
\log_{10}(2\theta_*) = (-0.2765\pm0.0093)(g-i)_o-0.2i_o+(0.7286\pm0.0076)  
\end{equation}
The data and best fit relations can be seen in Figure~\ref{fig:boyajian_new}. As expected, the $(r-i)_o$ relation is less accurate ($rms\sim0.05$) than the $(g-i)_o$ relation ($rms\sim0.04$), especially for the coolest stars with $(r-i)_o\ge1$ mag. But the accuracy is similar for a MOA-2019-BLG-009 source with $(r-i)_o\sim0.75$ mag. 
\section{Magnitude systems transformation}
\label{sec:mag_trans}

Below is the photometric transformation used in this work. MOA is calibrated to the OGLE-III catalog \citep{Udalski2003,Bond2017} using\footnote{http://www.massey.ac.nz/~iabond/staging/mb19008.o3c.dat}:
\begin{equation}
\label{eq:MOA_IOIII}
I_{\rm OGLE_{\rm III}}= R_{\rm MOA} + (28.176\pm0.002) - (0.1869\pm0.0008)(V_{\rm MOA}-R_{\rm MOA}) \pm 0.08 \end{equation}
and
\begin{equation}
V_{\rm OGLE_{\rm III}}= V_{\rm MOA} + (28.346\pm0.002) - (0.0888\pm0.0008)(V_{\rm MOA}-R_{\rm MOA})\pm 0.08
\end{equation}

\bibliographystyle{aasjournal}  % needs package natbib
\bibliography{KB19008.bib}

%\begin{thebibliography}{}

%\end{thebibliography}

%% This command is needed to show the entire author+affilation list when
%% the collaboration and author truncation commands are used.  It has to
%% go at the end of the manuscript.
%\allauthors

%% Include this line if you are using the \added, \replaced, \deleted
%% commands to see a summary list of all changes at the end of the article.
%\listofchanges

\end{document}